\documentclass[11pt,final]{article}
\pdfoutput=1
\usepackage{MRnotes}

\newcommand{\beq}{\begin{equation}}
\newcommand{\eeq}{\end{equation}}
 
\newcommand{\bea}{\begin{eqnarray}}
\newcommand{\ea}{\end{eqnarray}}

\title{How to tame your (black hole) saddles:\\ Lessons from the Lorentzian Gravitational Path Integral}

 \author{Maciej Kolanowski and Donald Marolf}
\affiliation{
	Department of Physics, University of California, Santa Barbara, CA 93106, USA}
\emailAdd{mkolanowski@ucsb.edu}
\emailAdd{marolf@ucsb.edu}

\abstract{We resolve a puzzle associated with the spherically-symmetric sector of the AdS$_4$ Einstein-Maxwell partition function with inverse temperature $\beta$. Since charge is quantized, the semiclassical limit of the partition function is expected to be given by a sum over complex black hole solutions obtained by shifting the associated chemical potential $\mu$ by $\frac{2\pi i n}{e \beta}$ in terms of the relevant charge quantum $e$.   However, the sum over all such saddles turns out to diverge at any finite value of $\beta$. We therefore consider a definition of this partition function as an integral over a space of metrics that are real and of Lorentz-signature up to the presence of certain conical singularities.  A Picard-Lefschetz analysis shows that only a finite subset of the above saddles contribute to our integral at finite $\beta$, and thus that the sum over such saddles converges.  The low temperature limit  is nonetheless associated with a convergent sum over all  saddles that (as $\beta \rightarrow \infty$) approach the usual large real Euclidean black holes.  We also analyze the analogous partition function for the (uncharged) BTZ black hole in the ensemble defined by fixing an angular velocity $\Omega$ up to shifts by 
$\frac{2\pi i m}{s \beta}$, where $s=\frac{1}{2}$ or $s=1$ depending on the presence of absence of fermionic states.  In this case, at all $\beta$ we find that all saddles contribute and that the sum over saddles converges.  We also comment briefly on the apparent lack of utility of the so-called KSW condition in our context.
}

\begin{document}\maketitle
\section{Introduction}

Gravitational partition functions are traditionally studied using Euclidean path integrals.  However,  outside of particularly simple two-dimensional theories \cite{Saad:2019lba}, the specific contour over which the integral is to be performed is  typically left unspecified.  In particular, let us recall that the integral over real Euclidean metrics is famously divergent due to the conformal factor problem.  As a result,   a meaningful Euclidean path integral must be associated with a nontrivial contour in the space of complex Euclidean metrics, and the path integral is specified only after such a contour is chosen \cite{Gibbons:1978ac}.  However, as recently reviewed in \cite{Horowitz:2025zpx}, there is at present no generally accepted recipe for  which contour is to be used.

Recall also that gravitational path integrals are typically studied in the semiclassical approximation. Despite classic investigations such as \cite{Halliwell:1988ik,Halliwell:1989dy,Halliwell:1989vu,Halliwell:1990tu}, issues regarding the choice of contour are sometimes seen as being  of little concern in this context.  While in principle the choice of contour determines which saddles are relevant to the path integral via Picard-Lefschetz theory, physically-sensible results have often been obtained in practice by simply finding interesting saddles and, perhaps, making perturbative arguments regarding their so-called `stability,'  
 see e.g.
\cite{Gibbons:1978ac,Garriga:1997wz,Gratton:1999ya,Kol:2006ga,Marolf:2022ntb, Liu:2023jvm}. For reasons that we explain further in section \ref{sec:chsing} below (see also appendix \ref{subsec:Euclidean}),  this has plausibly been possible due to a historical focus on real Euclidean saddles (perhaps after simple analytic continuations in the context of rotating black holes).  

However, one would expect such a seat-of-the-pants approach to be more difficult in contexts where saddles are intrinsically complex.  Indeed, over 175 years of complex analysis have shown that analytic continuation of saddles generally leads to Stokes' phenomenon \cite{Stokes:1847,Stokes:1858}, in which the coefficient with which a saddle contributes to the semiclassical expansion can change discontinuously when parameters are varied.  In such settings one expects to need a more complete specification of the desired path integral.

We investigate two path integrals below that are associated with grand canonical partition functions at complex values of the chemical potentials for electric charge  (in one case) and angular momentum (in the other case).  The relevant theories are the spherically-symmmetric sector of AdS$_4$ Einstein-Maxwell theory (associated with AdS Reissner-Nordstr\"om black holes) and pure Einstein-Hilbert gravity in AdS$_3$ (associated with BTZ black holes).  As emphasized in e.g. \cite{Heydeman:2020hhw}, the fact that these charges are quantized makes complex potentials intrinsically relevant to {\it any} gravitational partition function and, in both cases, leads to an infinite set of complex saddles.  In the former case, including all associated complex saddles turns out to yield divergent results at all finite inverse temperatures $\beta$.  

We therefore analyze the various saddles using an approach suggested in \cite{Marolf:2022ybi, Chen:2025leq}; see also \cite{Held:2024qcl,Held:2026huj,Held:2026bbo} for other applications.  This approach builds on the work of \cite{Marolf:2020rpm,Colin-Ellerin:2020mva,Colin-Ellerin:2021jev} to define the path integral using a contour associated with real {\it Lorentzian}-signature configurations that are allowed to contain certain classes of codimension-2 `conical' singularities (or `helical' singularities), with the action of such singular spacetimes defined as in \cite{Louko:1995jw} (see also \cite{Neiman:2013ap} for related work with different choices of sign), or by the charged and (in 2+1 dimensions) rotating generalization described in \cite{Chen:2025leq}.\footnote{While we expect further such generalizations to be possible, the  results of \cite{Chen:2025leq} suffice for the cases considered in this work.}  
In this framework, we find that  only a finite subset of the above saddles contribute at finite $\beta$, and thus that the sum over such saddles converges.  The limit $\beta \rightarrow \infty$ is nonetheless associated with a convergent sum over all saddles for which this limit approaches a familiar large AdS black hole with real Euclidean metric.

The remainder of this section will be  devoted to a more pedagogical review of the relevance of complex potentials to black holes in contexts with quantized charges (section \ref{sec:puzzle}) and to further discussion of why a Lorentzian approach is natural (section \ref{sec:Lor}).   Section \ref{sec:chsing} then provides a brief explanation of the path integrals proposed in \cite{Marolf:2022ybi,Chen:2025leq}, together with a discussion of their strategy for how such path integrals might be usefully approximated in the semiclassical limit.  The end result is then a simple and physically-natural expression  for the leading-order semiclassical partition function expressed (see \eqref{eq:Zapprox2}) as an integral over the space of stationary Lorentz-signature black holes weighted by $e^{A/4G}$, where $A$ is the corresponding horizon area. Section \ref{sec:PL} then performs a Picard-Lefschetz analysis of this integral for the partition function of AdS$_4$ Einstein-Maxwell theory in the spherically-symmetric sector.  The AdS$_3$ pure Einstein-Hilbert case is then studied in section \ref{sec:BTZ}.  We close with some final discussion in section \ref{sec:disc}, which in particular compares our approach with attempts to use the KSW condition \cite{Kontsevich:2021dmb,Witten:2021nzp} for similar purposes. 

\subsection{A puzzle}
\label{sec:puzzle}

We wish to compute the grand canonical partition function of the Einstein--Maxwell ($\textrm{AdS}$) theory at the temperature $\beta^{-1}$, chemical potential $\mu$ and angular velocity $\Omega$.  In the present work we will consider rotation only in 2+1 dimensions so that will suffice to treat both $\Omega$ and the corresponding angular momentum $J$ as scalars. In the semiclassical approximation, according to the standard lore of black hole thermodynamics, we would expect to find that the answer is given by $Z\approx^{-S_E}$, where $S_E$ is the Euclidean action of a Kerr-Newman black hole with appropriate potentials. 

However, following the standard Euclidean approach leads to the following problem. We would like to interpret the partition function as
\begin{equation}
    Z = \textrm{Tr} e^{-\beta \left( H - \mu Q - \Omega J \right)},
\end{equation}
where for concreteness one might use the AdS/CFT correspondence to think of the trace as being taken over the Hilbert space of some dual CFT.  Since charge and angular momentum are quantized, the expression on the right hand side is periodic in the imaginary directions of $\mu$ and $\Omega$; i.e., it is invariant under the independent shifts
\begin{subequations}
    \begin{equation}
        \mu \mapsto \mu + \frac{2\pi i n}{q \beta},
\label{eq:shiftm}    \end{equation}
    \begin{equation}
        \Omega \mapsto \Omega + \frac{2\pi i m}{s \beta},
\label{eq:shiftO}\end{equation}
    \label{eq:shiftm}
\end{subequations}
where $n, m \in \mathbb{Z}$, $q$ is the elementary charge and $s$ is the elementary spin (either $1/2$ or $1$, depending on the presence of fermions in the theory).
In contrast, when it is evaluated on a single black hole solution, the expression $e^{-S_E}$ is clearly not periodic in this sense.

As emphasized in \cite{Heydeman:2020hhw}, the problem is resolved by noting that there is more than one saddle that satisfies the boundary conditions given by any fixed values of $\mu$ and $\Omega$. Indeed, since the associated gauge groups are compact,  the shifts \eqref{eq:shiftm}, \eqref{eq:shiftO} in $\mu$ and $\Omega$  can be accomplished by acting with large gauge transformations near infinity.   As a result, given any bulk configuration with any complex potentials $\mu+\frac{2\pi i n}{q \beta},\Omega+\frac{2\pi i n}{s \beta}$, we can use the above large gauge transformation to construct a solution having the same action\footnote{Here we refer to the action with boundary terms appropriate to the desired grand canonical partition function.} and satisfying the boundary conditions defined by the desired desired potentials $\mu, \Omega$ (and satisfying the same fermion boundary conditions for $s=1/2$). With this understanding the desired path integrals are manifestly invariant under the shifts \eqref{eq:shiftm} and \eqref{eq:shiftO}.

Nevertheless, it is traditional to describe black hole configurations using a global trivialization of the gauge bundle. I.e., we take the entire spacetime to be described by a single coordinate chart, and we take the metric and gauge field components to be smooth functions in this chart.   Doing so associates each black hole solution with a {\it unique} value of $\mu, \Omega.$  But since the boundary conditions in fact specify only equivalence classes of $\mu, \Omega$ under \eqref{eq:shiftm}, \eqref{eq:shiftO}, configurations corresponding to all values of $n,m$ in 
\eqref{eq:shiftm}, \eqref{eq:shiftO} must contribute to the associated path integral.  In other words, with this convention our path integral must include an explicit sum over $n,m\in {\mathbb Z}$.


In certain situations this sum can be done exactly.  For example, let us consider Jackiw-Teitelboim (JT) gravity  coupled to $U(1)$ and $SU(2)$ gauge fields (describing Maxwell field and rotations, respectively). In this case, the theory is one-loop exact.  Furthermore, for each shifted potential one can find a new saddle (and its one-loop contribution), and one can then sum over such saddles using Poisson-resummation formula \cite{Heydeman:2020hhw}. In particular, the sum is given by a convergent series.  Since the sum is manifestly periodic, it beautifully leads to the density of states with only discrete charges for the gauge fields. 

One could then try to do the same in four-dimensions (up to the lack of one-loop exactness). It is not hard to write down all Kerr-Newman solutions with prescribed boundary conditions. However, as will become explicit in section \ref{sec:PL}, one immediately finds the sum over such saddles to diverge. In particular, the (real part of the) on-shell actions associated with this infinite class of saddles are not bounded from below \cite{Harlow2022}. It thus seems that some argument is needed to remove the infinite number of saddles.   As stated in the introduction, we will provide such an argument  below starting from the Lorentzian Path Integral.

\subsection{Why Lorentzian?}
\label{sec:Lor}

Before describing the details of our construction, it will be beneficial to explain why we expect the Lorentzian Path Integral to be a good guiding principle (despite the fact that the problem of current interest may seem to be more naturally Euclidean in nature). Let us start by recalling that, to properly evaluate the stationary phase approximation to any integral -- even in the finite-dimensional case -- it is not enough to know all the saddles of the integrand. We also need to know if our contour of integration can be deformed to run appropriately through each saddle. This requires in particular that one knows which contour (or which equivalence class of contours) is to be used to define the desired integral in the first place.  

However, it is well-known that one cannot simply define the gravitational path integral as an integral over all real Euclidean metrics.  A primary obstacle for any gravitational system arises from the conformal factor problem, which renders the Euclidean action unbounded from below on the space of real Euclidean metrics \cite{Gibbons:1978ac}.  An integral over real Euclidean metrics must thus diverge. 

    In the linearized theory about simple real Euclidean saddles, physically acceptable results are often obtained by simply rotating by hand the contour of integration for 'pure trace' modes of the metric as suggested in \cite{Gibbons:1978ac}, generalizing this idea as in \cite{Marolf:2022ntb, Liu:2023jvm}, or using methods that impose constraints; see e.g. \cite{Garriga:1997wz,Gratton:1999ya,Kol:2006ga}.\footnote{Although they are typically not described in this way, such methods thus effectively Wick-rotate Euclidean-signature Lagrange multipliers.} However, as reviewed recently in \cite{Horowitz:2025zpx} and \cite{Held:2026huj}, there is little understanding of such ideas at the non-perturbative level required to determine the relevance of general complex saddles.


In addition, a further issue in the current case is that boundary conditions associated with complex $\Omega, \mu$ prohibit consideration of field configurations that are even asymptotic to those that are real and Euclidean. 
As a result, whenever $\Omega, \mu$ have non-zero imaginary parts, we cannot think of the desired configurations as being perturbatively close to real Euclidean configurations.  It becomes unclear how one might even attempt to apply the recipes above  that have often worked at the linearized level.

For example, in the static sector (with $\Omega = J = 0$), it is natural to think about the Euclidean Maxwell field as having imaginary Euclidean-time component $A_\tau$ whose asymptotic behavior is controlled by $A_\tau \sim i\mu$. However, if $\mu$ is complex, so must be $A_\tau$, even asymptotically. As a result, we cannot consistently formulate the resulting path integral as an integral over either purely real or purely imaginary values of $A_\tau$ while simultaneously satisfying the required boundary conditions.

Now, it has long been argued that issues with the Euclidean path integral should be resolved by instead using path integrals that are fundamentally defined as integrals over real Lorentz-signature metrics; see e.g. \cite{Hartle:2020glw,Schleich:1987fm,Mazur:1989by,Giddings:1989ny,Giddings:1990yj,Marolf:1996gb,Gratton:1999ya,Dasgupta:2001ue,Ambjorn:2002gr,Feldbrugge:2017kzv,Feldbrugge:2017fcc,Feldbrugge:2017mbc,Brown:2017wpl}.\footnote{A primary motivation for this approach is that, at least for appropriate spacetimes with topology $\Sigma \times {\mathbb R}$, the Lorentzian path integral can be derived from canonical quantization of the theory.  It should thus be free of divergences after imposing an appropriate UV cutoff, and should therefore resolve at least a large part of the conformal factor problem.}  Note that the contour of real Lorentz-signature metrics can also be thought of as a particular choice of contour in the space of complexified Euclidean metrics.  Indeed, one might say that the main point of \cite{Colin-Ellerin:2020mva,Colin-Ellerin:2021jev}
    was to argue that, so long as one allows certain codimension-2 singularities in the Lorentzian structure, the statement that one may think of Lorentzian metics in terms of complex Euclidean metrics remains true on manifolds of general\footnote{It would be useful to rigorously establish precisely how general is the relevant class of topologies in arbitrary dimensions.} topology which need not admit smooth real Lorenz-signature metrics at all.  See also comments about such singularities in \cite{Dong:2016hjy,Marolf:2020rpm}.

Furthermore, despite the issues noted above with Euclidean boundary conditions in sectors with non-trivial $n,m$, 
we will see that the desired partition functions can nevertheless be formulated using only integrals over such real Lorentz-signature contours.
Indeed, in the Lorentzian context the path integral more naturally computes matrix elements of an operator that we may write in the form:
\begin{equation}
    U_T = e^{-i T \left(
    H -\mu Q - \Omega J
    \right)}.
    \label{eq:UT}
\end{equation}
In particular, for an asymptotically AdS system with a CFT dual, there is a corresponding unitary operator in the CFT.

Following \cite{Marolf:2022ybi}, we can then recover our original partition function by computing
\begin{subequations}
    \begin{equation}
    \label{eq:fTtransform}
Z(\beta, \mu, \Omega):=    \textrm{Tr} e^{-\beta (H-\mu Q - \Omega J)} = \textrm{Tr} \int_\mathbb{R} \mathrm{d}T f_\beta(T) U_T,
\end{equation}
where
\begin{equation}
    f_\beta(T) = -\frac{e^{iT E_0 - \beta E_0}}{2\pi i (T+i \beta)},
\end{equation}
\end{subequations}
and where $E_0$ is any number below the infimum of the spectrum\footnote{This construction applies only when this operator is bounded below.} of $H - \mu Q - \Omega J$. Notice that $U_T$, being an unitary operator acting on an infinite-dimensional Hilbert space, is not trace-class. As such, the trace and the integral in the expression above do not commute.

Due to charge and angular momentum quantization, the operator $U_T$ defined by \eqref{eq:UT} are periodic as a function of $\mu$ and $\Omega$, but this time in the real direction:
\begin{subequations}
    \begin{equation}
    \label{eq:realmushif}
        \mu \mapsto \mu + \frac{2\pi n}{q T},
    \end{equation}
    \begin{equation}
    \label{eq:realomegashiftt}
        \Omega \mapsto \Omega + \frac{2\pi  m}{s T}.
    \end{equation}
\end{subequations}
As a result, the corresponding path integral can again be written as an integral over $T$ of a sum over sectors with shifted values of $\mu, \Omega$ (and with $T$-dependent shifts). The parameter $\mu$ still controls the asymptotic behavior of $A_t$.  But in Lorentz signature, we have asymptotically $A_t \sim \mu$, so it is easy to accommodate real shifts in $\mu$. Analogous statements can also be made for $\Omega$ and the off-diagonal $g_{t\phi}$ term in the metric. While it remains to describe other aspects of the relevant gravitational path integrals, we will do so in section \ref{sec:chsing} in direct analogy with \cite{Marolf:2022ybi} and \cite{Chen:2025leq}. In summary then, a contour that integrates over asymptotically real Lorentzian configurations can be used to consistently implement all relevant boundary conditions, though  analogous statements fail for the (asymptotically) Euclidean case.


One might also ask if we could instead use a Euclidean gravitational path integral to compute the quantity\footnote{Again, expression \eqref{eq:euclidean_trace} might be most clearly understood as referring to a description in terms of a dual CFT.} 
\begin{equation}
    \textrm{Tr} e^{-\beta (H - i \mu Q - i \Omega J)} \label{eq:euclidean_trace}
\end{equation}
for real $\mu, \Omega$, and if we could then analytically continue the result to imaginary values of $\mu,\Omega$.   If one could somehow deal with the conformal factor problem, then such an approach would indeed allow for imposing purely real Euclidean boundary conditions at infinity.  However, since $Q$ and $J$ are not bounded from below (in contrast to the combination $H-\mu Q-\Omega J$ for small $\mu, \Omega$), one may expect that the desired analytic continuation would involve further subtleties. In particular, an integral transform analogous to  \eqref{eq:fTtransform}, but which integrates over $\mu$ and/or $\Omega$ instead of $T$, will not give the desired result (since the contour can no longer be closed in the lower half plane).   Moreover, while at finite Newton's constant $G$ the partition function might expected to be an analytic function of $\mu$ and $\Omega$, this is certainly {\it not} the case once we have taken the semiclassical limit.  This can be seen directly from the fact that that limit exhibits phase transitions (which are just direct analogs of the well-known Hawking-Page transition \cite{Hawking:1982dh}). Nevertheless, for completeness we will use the Lorentzian approach to determine which saddles contribute to \eqref{eq:euclidean_trace} in appendix \ref{subsec:Euclidean}.

\section{Integrating over metrics with conical and helical singularities}
\label{sec:chsing}

It remains to explain how the partition function \eqref{eq:fTtransform} is to be described using Lorentzian gravitational path integrals. 
To motivate the proposal of \cite{Marolf:2022ybi} and the generalization in \cite{Chen:2025leq}, let us first note that \eqref{eq:fTtransform} contains a trace.  As a result, it is natural to expect a corresponding (say, asymptotically AdS) bulk gravitational path integral to sum over spacetimes whose conformal boundary ${\cal B}$ has no boundary ($\partial {\cal B} = \emptyset$).

We may then also note that 
\eqref{eq:fTtransform}
refers to a one-parameter family of operators $U_T$ that enact time-translations by different amounts $T$ but which are generated by the same Hamiltonian $H$.    It is then natural to associate each $U_T$ with a boundary spacetime of topology $S^1 \times \Sigma$ (where an a priori factor of the interval has been compactified to $S^1$ due to the above-mentioned trace).  We assume each boundary spacetime to have a Killing field that generates translations along the $S^1$ that are periodic with period $|T|$.  We also assume that the metrics on boundary spacetimes with different values of $T$ are locally identical, and that the same normalization is used to define the magnitude of Killing translations on each spacetime.   In this sense, our boundary spacetimes differ only by the choice of period for the $S^1$.

The proposal of \cite{Marolf:2022ybi} was thus to define the partition function \eqref{eq:fTtransform} with
an explicit integral over $T$ in parallel with  \eqref{eq:fTtransform}.  For each real $T$ one then integrates 
over a class of real Lorentz-signature spacetimes satisfying the above boundary conditions.   For positive values of $T$ the integrand is taken to be $e^{iS}$, while for negative values of $T$ it is taken to be the complex conjugate $(e^{iS})^*$ (and thus to be $e^{-iS}$ when $S$ is real).\footnote{In fact, motivated by the assumption that one should sum over both signs of Lagrange multipliers in the canonical formalism, we believe that the path integral should sum over assignments of $e^{iS}$ vs. $(e^{iS})^*$ to all possible spacetime regions, but with such assignments constrained by boundary conditions.  Here we simply require $e^{iS}$ ($[e^{iS}]^*$) in any neighborhood of a boundary with $T>0$ ($T<0$) and we simplify the discussion by treating this choice as being uniform over any given spacetime.}  The latter convention reflects the expectation that, to the extent that either quantity is defined, we should have $Tr\, U_T = (Tr\, U_{-T})^*$.

\begin{figure}[!h]
    \centering
\includegraphics[width=0.3\linewidth]{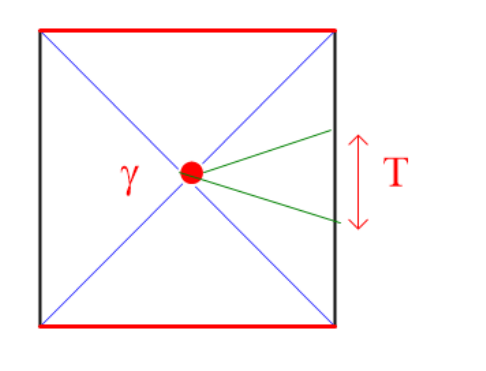}
    \caption{The conformal diagram of an AdS Schwarzschild black hole is shown with two marked surfaces differing by a time translation of magnitude $T$.  Cutting along these surfaces defines a wedge.  Identifying the two cut-surfaces of the wedge then yields a time-periodic spacetime with what we may call a Lorentzian conical singularity (red dot). }
    \label{fig:BlackHoleQuotient}
\end{figure}

An important point, however, is that one might wish to include bulk spacetimes with topology $D^2\times \Sigma$, where $D^2$ is the usual two-dimensional disk.  But there are no smooth Lorentz-signature metrics with this topology that satisfy the above boundary conditions.  Instead, one finds a codimension-2 surface at which the causal structure becomes singular. Indeed, a prototypical example is given by considering a stationary black hole spacetime with bifurcate horizon, slicing it open from the bifurcation surface to infinity along two surfaces related by a horizon-generating Killing field translation of magnitude $T$, keeping only the finite wedge that lies between these cuts, and finally identifying the two cuts to create a time-periodic boundary; see figure \ref{fig:BlackHoleQuotient}.   

One must thus take the domain of integration to include metrics with this sort of singularity (and perhaps with others as well).  In particular, one must take care to define the gravitational action on such spacetimes.  As previously advertised, this is done following the treatment of the analogous two-dimensional case in \cite{Louko:1995jw}.  The higher-dimensional case was argued in \cite{Colin-Ellerin:2020mva} to be essentially the same.

For Einstein-Hilbert gravity, the associated result may be written as in \cite{Marolf:2022ybi} in the convenient form
\begin{eqnarray}
\label{eq:defectSEH}
S_{EH} &=& \frac{1}{16\pi G_N} \int_M \sqrt{-g}R : = \lim_{\epsilon \rightarrow 0} \left( \frac{1}{16\pi G_N} \int_{M\setminus U_\epsilon} \sqrt{-g}R  - \frac{1}{8\pi G_N} {\cal P} \int_{\partial U_\epsilon} \sqrt{|h|}K  \right) \\ &+& i \left(\frac{\cal N}{4}-1\right)\frac{A_\gamma}{4G_N}.
\end{eqnarray}
Here $U_\epsilon$ is a small neighborhood containing the singular codimension-2 surface $\gamma$, $A_\gamma$ is the area of $\gamma$, $G_N$ is Newton's constant, $\cal P$ denotes the principle part of the associated integral (whose integrand has simple poles where $\partial U_\epsilon$ becomes null), and ${\cal N}$ denotes the number of orthogonal null geodesics to $\gamma$ at each point of $\gamma$.  In this notation,  Minkowski space (with no singularities) has ${\cal N}=4$ while the example in figure \ref{fig:BlackHoleQuotient} has ${\cal N}=0$.   See \cite{Colin-Ellerin:2020mva} and \cite{Marolf:2022ybi} for further discussion.  In contrast, the action for minimally-coupled matter is given by just integrating the standard Lagrangian density over the spacetime with no special contributions from $\gamma$.  
While there is certainly more to understand regarding theories with higher derivative terms, and while it will not be relevant to our discussion below, we mention that in such cases it was proposed in \cite{Colin-Ellerin:2020mva} to define the action using the Legendre transform of the action described in appendix B of \cite{Dong:2019piw}.
 
The attentive reader will notice that the $A_\gamma$ term at the end of  \eqref{eq:defectSEH} is imaginary.  This imaginary contribution arises from the use of a complex regulator in computing the action \eqref{eq:defectSEH}.  In particular, the computation is performed by approximating the desired singular spacetime using a smooth complex metric.  The use of a complex metric is essential since, as stated above, there is no smooth real Lorentz-signature metric that satisfies the desired boundary conditions.\footnote{The sign of the imaginary regulator was chosen in \cite{Louko:1995jw} by requiring standard matter path integrals to converge.  This is the same idea that motivates what is often called the Kontsevich-Segal-Witten criterion \cite{Kontsevich:2021dmb,Witten:2021nzp} (though the origins of this idea date back at least to \cite{Halliwell:1989dy}).  The sign of this regulator then determines the sign of the imaginary term in \eqref{eq:defectSEH}.  The use of convergence of the matter part of the path integral in determining this sign is consistent with our rule that for negative $T$ the integrand is $(e^{iS})^*$.}

Specifying the rest of the form of the path integral proposed in \cite{Marolf:2022ybi} now involves tying up several loose ends.  Let us first note that the above imaginary part means that metrics with a Lorentzian conical singularity with ${\cal N}=0$ cause the integrand to increase in magnitude exponentially with $A_\gamma$.  One might be concerned about the convergence of such integrals, though the fact that the integrand oscillates in other directions might potentially provide sufficient cancellations to bring it under control.  In order to maximize the chance of this occurring, \cite{Marolf:2022ybi} proposed that the integral over $A_\gamma$ be performed last and, in particular, that it be performed only {\it after} performing the integral over $T$.  In other words, the proposal was to define the desired partition function to be

\begin{equation}
\label{eq:Zdef}
Z(\beta, \mu, \Omega) : = 
\int dA_\gamma \, dT f_\beta(T)
Z_T(\mu, \Omega;A_\gamma), 
\end{equation}
where $Z_T(\mu, \Omega;A_\gamma)$ is the path integral over all real Lorentzian field configurations satisfying the above fixed-$T$ boundary conditions, with $\mu, \Omega$ specifying the asymptotic form of the usual (here, real!) components of the metric and gauge field,  and with the total area of all (spacelike) Lorentzian conical singularities fixed to $A_\gamma$. After taking \eqref{eq:Zdef} as a definition, it is then possible in certain cases that one will be allowed to exchange various orders of integration without changing the final result; see \cite{Held:2026huj} for further discussion of this point.

Another important ingredient in the approach of  \cite{Marolf:2022ybi} was to recall from \cite{Colin-Ellerin:2020mva} that 
spacetimes constructed in analogy with figure \ref{fig:BlackHoleQuotient} (by cutting wedges out of stationary Lorentzian black hole spacetimes and gluing the cut edges together) are in fact saddle-points of the constrained path integral that computes each  $Z_T(\mu, \Omega;A_\gamma)$.  Indeed, holding fixed $A_\gamma$ precisely allows the presence of a (here, complex) delta-function Ricci curvature associated with the Lorentzian conical singularity so long as the surface $\gamma$ at which it is located is an extremal surface in the corresponding smooth stationary black hole spacetime.\footnote{See appendix A of \cite{Dong:2019piw} for a discussion of the extremality condition in Euclidean signature.}  Furthermore, 
the fact that such saddles lie on the defining contour of integration allowed \cite{Marolf:2022ybi}
to show that, for any real $T$, such saddles will necessarily contribute to 
$Z_T(\mu, \Omega;A_\gamma)$ in the associated Picard-Lefschetz analysis.  This then motivated the suggestion of \cite{Marolf:2022ybi} that 
one might usefully approximate 
\eqref{eq:Zdef} by replacing 
$Z_T(\mu, \Omega;A_\gamma)$ with $e^{iS}$ (or $[e^{iS}]$) evaluated on saddles of this form.\footnote{\label{foot:caveats} This suggestion assumes that the above saddle not only contributes at each real $T$, but also that it always dominates.  It also assumes that subdominant terms will remain negligible in comparison after performing the integral over $T$.  As we will soon see, this is equivalent to assuming that no Stokes' phenomenon causes the saddle to become irrelevant under an analytic continuation $T \rightarrow -i\beta$.}.

In the current context, for consistency with charge quantization one should at this stage include all saddles associated with non-trivial values of $n,m$ in \eqref{eq:realmushif} and \eqref{eq:realomegashiftt} since such shifted saddles are again Lorentz-signature and real.  In particular, on the $(n,m)$ saddle the potential difference\footnote{\label{foot:dPhi} Here $\Delta \Phi = \int F_{ab}\xi^a t^b$ where $\xi^a$ is the horizon-generating Killing field in the original black hole spacetime and $t^b$ is the tangent to the path over which the integral is performed.} $\Delta \Phi$ between the codimension-2 singularity and infinity takes the value $\mu+\frac{2\pi n}{qT}$, and the analogous construction for angular velocity (say, defined by the gauge field that emerges from Kaluza-Klein reduction along the direction of rotation) gives $\Delta \Omega = \Omega+\frac{2\pi m}{sT}$.

Since our saddles have a Killing field, using the canonical version of the Einstein-Hilbert action and taking the Killing translation to act in the time direction, one quickly finds the relevant action to be
\begin{eqnarray}
S &=& -T(E-\left[\mu + \frac{2\pi n}{qT}\right] Q - 
\left[\Omega + \frac{2\pi n}{sT}\right]J) - i \frac{A_\gamma}{4G_N} \\ &=&
-T(E-\mu Q - \Omega J) - \frac{2\pi n}{q} Q - 
\frac{2\pi n}{s}J- i \frac{A_\gamma}{4G_N}, 
\end{eqnarray}
where we have now explicitly indicated the effect of the shifts \eqref{eq:realmushif} and \eqref{eq:realomegashiftt}, and where $E,Q,J$ are determined by the data $\mu,\Omega, n, m, A_\gamma$ according to the allowed values for real stationary Lorentz-signature bifurcate black holes.

Assuming $E-\mu Q-\Omega J$ to be bounded below on the space of real stationary bifurcate Lorentz-signature hole geometries, 
replacing $Z_T(\mu, \Omega;A_\gamma)$ in \eqref{eq:Zdef} with $e^{iS}$ (or $[e^{iS}]$)  allows one to close the contour of integration for $T$ in the lower half plane.  Cauchy's theorem then computes the $T$-integral in terms of the residue at $T=-i\beta$ to yield
\begin{equation}
\label{eq:Zapprox}
Z(\beta, \mu, \Omega) : \approx 
\sum_{n,m} \int dA_\gamma \ 
e^{-\beta(E-\mu  Q-\Omega J)}e^{-\frac{2\pi n}{q}iQ} e^{-\frac{2\pi m}{s}iJ} e^\frac{A_\gamma}{4G_N}. 
\end{equation}
One can study the resulting $A_\gamma$ integral in more detail either by  performing the integral numerically or by carefully applying the saddle-point approximation.  

In particular, since what remains is an integral over only one variable, it becomes tractable to perform a complete Picard-Lefschetz analysis to determine the relevance of any complex saddles.  Here we remind the reader that a saddle is relevant in the Picard-Lefschetz sense \cite{FAs,FP,AGV,BH,BH2,H} when the steepest {\it ascent} contour from the saddle has non-zero intersection number with the contour of integration, and that the saddle is irrelevant in this sense when this intersection number vanishes; see e.g. \cite{Witten:2010cx} for a physicist-oriented review, and see appendix A of \cite{Held:2026huj} for an even more condensed review that nevertheless explicitly discusses contours with finite endpoints.    

We emphasize that saddles of the $A_\gamma$ integral in \eqref{eq:Zapprox} are generally associated with saddles of the unconstrained version of our Lorentzian path integral and, in particular, with saddles where time is periodic with period $T=-i\beta$.  Since $\beta$ is real, these may be either real saddles of Euclidean signature or more general complex spacetimes.  In either case, in these spacetimes the electric potential difference between the horizon and infinity will take the complex values $\Delta \Phi = \mu + \frac{2\pi n}{q\beta}i$, and similarly $\Delta \Omega= \Omega+\frac{2\pi m}{s\beta} i$.  Here we use the same conventions for $\Delta \Phi, \Delta \Omega$ as above (see again footnote \ref{foot:dPhi}) and, in particular, we continue to take the norm of $\xi^a$ to be negative; i.e., we continue to define $\Delta \Phi$ by using the `Lorentz-signature' Killing field $\xi^a$ and Maxwell field $F_{ab}$ even for `Euclidean' saddles.  Indeed, here we have merely set $T=-i\beta$ in the expressions for $\Delta \Phi, \Delta \Omega$ that were given above.

Note  that  an attempt to sum over $n,m$ before integrating over $A_\gamma$ would appear problematic.  In particular, the sum over $n,m$ would explicitly build a pair of so-called Dirac combs (periodic arrays of Dirac delta-functions) that
would enforce quantization of {\it both} $Q$ and $J$.  However, $Q$ and $J$ in \eqref{eq:Zapprox} are both functions of $A_\gamma, \mu, \Omega$ determined by the space of real stationary Lorentz-signature black holes.  Generically one will thus find that, for given values of $\mu, \Omega$, there are simply no values of $A_\gamma$ for which the relevant $Q, J$ are both quantized in terms of the appropriate units.  At our current level of approximation, we thus find no saddle-point contributions at all to the desired partition function.

This, however, should not be a surprise. The issue is simply that the above Dirac combs have structure fine enough to be sensitive to subleading corrections to the stationary phase approximation used to replace $Z_T(\mu, \Omega;A_\gamma)$ by $e^{iS}$ (or $[e^{iS}]^*$).  Indeed, one can see that restricting to configurations with quantized charges should actually have rather little effect since the variance of either $Q$ or $J$ along a steepest descent contour through a semiclassical peak will generally be much larger than the associated charge quantum.

It is less immediately clear whether one finds the same subtlety if one integrates over $A_\gamma$ before performing the sum over $n,m$. 
Nevertheless, this discussion suggests that it would be useful to refine the above use of the stationary phase approximation, perhaps by identifying a small set $\nu$ of additional interesting degrees of freedom that can be treated in parallel with $A_\gamma$; i.e., for which we can first consider constrained saddles with fixed values of the additional parameters $\nu$ and then later explicitly perform the integral over $\nu$.  

Indeed, it was already suggested in \cite{Marolf:2022ybi} that this would be useful in the context of charged and rotating black holes and, in particular, that it might give additional insight into which saddles actually contribute to the desired partition function (see again comments in footnote \ref{foot:caveats}).  It was also suggested that it should be possible to choose the parameters $\nu$ so that, when restricted to stationary black hole solutions, the parameters would coincide with $Q$ and $J$.  One would then expect to be able to write the desired partition function in the form
\begin{equation}
\label{eq:Zapprox2}
Z(\beta, \mu, \Omega) : \approx 
\sum_{n,m} \int dA_\gamma\, dQ \, dJ \ 
e^{-\beta(E-\mu  Q-\Omega J)}e^{-\frac{2\pi n}{q}iQ} e^{-\frac{2\pi m}{s}iJ} e^\frac{A_\gamma}{4G_N}, 
\end{equation}
where the allowed values of $A_\gamma, Q, J$ are again determined by the space of real stationary Lorentz-signature black holes (regardless of whether those black holes have electric potential $\mu$ or angular velocity $\Omega$).  Here $E$ would then represent the energy of such a black hole with area $A_\gamma$, charge $Q$, and angular momentum $J$. 

It is worth emphasizing that expression \eqref{eq:Zapprox2} is precisely what one would expect from the idea that each stationary classical black hole solution is associated with $e^{A/4G_N}$ microstates.  Equation \eqref{eq:Zapprox2} might thus be reasonably taken as a starting point for a careful Picard-Lefschetz analysis even without motivating this result from the use of Lorentz-signature path integrals\footnote{Indeed, a reader who finds themselves skeptical of our Lorentzian starting point may choose to take this point of view. This perspective was recently used in \cite{Mahajan:2025bzo, Singhi:2025rfy, Ailiga:2025osa}  to argue against the use of certain complex black hole saddles, and to also argue against the use of naked singularities, either of which would otherwise destroy the standard Hawking-Page transition.}.  

Let us now note  that 
the integrand in \eqref{eq:Zapprox2} becomes real for $n=m=0$.  In this case, the contributing saddles are precisely those with real values of $A_\gamma, Q, J$ that yield  local maxima of the integrand (local minima of the corresponding action).\footnote{\label{foot:realint} This result relies on making certain choices for the non-perturbative corrections to these saddles as, if the integrant also has local minima, they  will necessarily lie on Stokes' rays.  Nevertheless, such choices can always be made.}  This may explain why perturbative approaches have tended to yield physically-reasonable results in past studies of real Euclidean saddles; see e.g.
\cite{Gibbons:1978ac,Garriga:1997wz,Prestidge:1999uq,Gratton:1999ya,Kol:2006ga,Marolf:2022ntb, Liu:2023jvm}.

For $U(1)$ Maxwell fields, and for angular momentum in 2+1 directions, the desired generalization \eqref{eq:Zapprox2} was derived in \cite{Chen:2025leq}.    In the Maxwell case, one simply fixes the electric flux $Q$ through the singular surface $\gamma$ and finds, under such a constraint, that stationary points of the fixed-flux Maxwell action are then allowed to have non-trivial holonomies around $\gamma$ even for infinitesimal loops.  As a result, at fixed $Q$,
we are now allowed to take a general stationary black hole solution and to generate a new constrained saddle by shifting its electric potential by an arbitrary real number; i.e., there is no need to respect any quantization condition. In this way we can generate a constrained saddle with arbitrary $Q$ and which nevertheless satisfies the boundary conditions associated with any given value of $\mu$, so that the integral in \eqref{eq:Zapprox2} does indeed run over all real values of $Q$ that are associated with stationary black hole solutions.  An analogous (but more complicated) construction was also shown to go through for angular momentum in 2+1 dimensions.  In both cases, repeating the reasoning of \cite{Marolf:2022ybi}  then leads to \eqref{eq:Zapprox2} as desired.  These cases will suffice for the analyses to be performed in the work below.
For future reference,  we also mention that one can of course think of \eqref{eq:Zapprox2} as giving $\mu, \Omega$ finite imaginary parts $-\frac{2\pi n}{q\beta}$, $-\frac{2\pi m}{s\beta}$, and that the arguments used to derive \eqref{eq:Zapprox2} hold equally well for non-integer values of $n,m.$ As such, the above results can be generalized to derive \eqref{eq:Zapprox2} for arbitrary complex values of $\mu, \Omega$ so long as the associated partition function is finite; see also the related discussion in appendix \ref{subsec:Euclidean}.

\begin{figure}[!h]
    \centering
\includegraphics[width=0.3\linewidth]{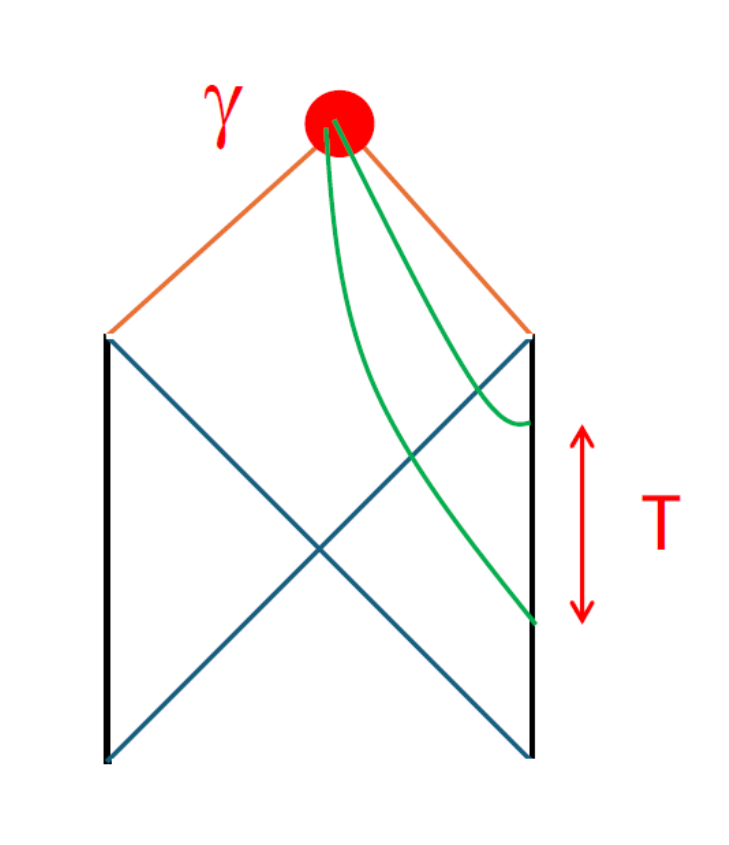}
    \caption{A conformal diagram of an AdS Reissner-Nordstr\"om black hole.  The two green surfaces stretch from the inner horizon $\gamma$ to the right boundary.  They are chosen to be related by a time translation of magnitude $T$.  Cutting along these surfaces, and identifying the two surfaces yields a time-periodic spacetime with a Lorentzian conical singularity at $\gamma$.  The periodic direction becomes null at the outer horizon and spacelike inside.}
    \label{fig:inner}
\end{figure}

A final fine point to discuss is whether, for fixed $Q,J$, the range of $A_\gamma$ in \eqref{eq:Zapprox2}  should include only allowed outer-horizon areas, or whether this range should also include corresponding inner-horizon areas.\footnote{We thank Harshit Rajgadia for emphasizing the importance of this question.}  It turns out that the recipes of \cite{Marolf:2022ybi} and \cite{Chen:2025leq} apply just as well in the inner horizon case, so it is natural to include them at the present level of analysis. In particular, the regularization of the conical deficits of \cite{Louko:1995jw} still holds and predicts the same sign for the imaginary part of the Lorentzian action.   Figure \ref{fig:inner} shows the inner horizon analogue\footnote{At least in some cases, the resulting spacetimes will feature closed null geodesics along the would-be outer horizon.  It would be interesting to better understand what effects, if any, this feature might have at the one-loop level.} of the construction performed in figure \ref{fig:BlackHoleQuotient}.  In our work below, we will thus include values of $(A_\gamma, Q, J)$ associated with inner horizons.

\section{AdS$_4$ Einstein-Maxwell theory with spherical symmetry}
\label{sec:PL}

We now study expression \eqref{eq:Zapprox2} for the AdS$_4$ Einstein-Maxwell partition function in the spherically symmetric sector. We will write this expression in the form
\begin{equation}
\label{eq:Zsum}
Z(\beta, \mu) = \sum_{n \in \mathbb{Z}} Z_{n},
\end{equation}
with
\begin{equation}
Z_{n}:= \int_{A>0} dA dQ \, e^{A/4}e^{-\beta  \left(E - \left[\mu + \frac{2\pi i n}{q\beta} \right] Q\right)} 
= \int_{A>0} dA dQ \, e^{A/4}e^{-\beta E} e^{\left( \beta \mu   - \frac{2\pi i}{q} n\right) Q},
    \label{eq:Znm}
\end{equation}
where spherical symmetry has imposed $J=0$ and $\Omega=0$ in   \eqref{eq:Zapprox2}.
We remind the reader that $E$ is a function $E = E(A,Q)$ of $A,Q$ determined by the energy of stationary real Lorentz-signature black holes with bifurcate horizons. In particular if, as suggested in section \ref{sec:chsing}, we allow $A$ to be the area of either an inner or outer horizon for an AdS Reissner-Nordstr\"om black hole, the domain of integration includes all $(A,Q) \in {\mathbb R^+} \times {\mathbb R}$; i.e.,  the condition $A>0$ is the only restriction on the domain.\footnote{Since extreme black holes are a set of measure zero, at the current level of approximation it does no harm to include them despite their lack of a bifurcation surface.  We may, however, anticipate that they will merit special treatment at the one-loop level.}  
For simplicity we have also set $G_N=1$  Finally, we remind the reader that the above expression is valid  only at the leading semiclassical level.  As a result,  the details of the correct measure have not been fully incorporated into \eqref{eq:Znm}.  Our leading-order analysis below will thus correspondingly assume only that the actual measure is a holomorphic function of $A,Q$.

 Let us note that (for real $\beta, \mu$), the integrand depends on $n$ only through a phase.  As a result, each $Z_{n}$ satisfies
\begin{equation}
    |Z_{n}| \le \int dA dQ |e^{A/4}e^{-\beta  \left(E - \left(\mu + \frac{2\pi i n}{q\beta} \right)Q\right) }| = Z_{0}.
    \label{eq:Z00}
\end{equation}
It is thus manifest that saddles with $n \neq 0$ are forbidden from contributing to our partition function when their Euclidean action $S_E$ satisfyies $|e^{-S_E}|> Z_{0}$.\footnote{In particular, since in the semiclassical approximation $Z_0$ can be no smaller than the maximum of the integrand in \eqref{eq:Z00}, ascent contours from such saddles cannot intersect the contour of integration.  This means that they cannot be relevant.   We again refer the reader to \cite{Witten:2010cx,Held:2026huj} for brief reviews of the conditions under which saddles are relevant.  In particular, explicit discussion of saddles for contours with finite endpoints can be found in appendix A of \cite{Held:2026huj}.}  

Rather than use the area $A$ as an integration variable, it is more convenient to integrate over radii $\mathcal{R} = \sqrt{\frac{A}{4\pi}}$. The energy of our constrained saddles is then
\begin{equation}
    E(Q, \mathcal{R}) = \frac{\mathcal{R}^3}{2L^2} + \frac{\mathcal{R}}{2} + \frac{Q^2}{2 \mathcal{R}},
    \label{eq:EQR}
\end{equation}
where $L$ is the usual AdS scale.  Let us also define $\mu_n := \mu + \frac{2\pi i n}{q\beta}$. Substituting \eqref{eq:EQR} into \eqref{eq:Z00}, the integral over $Q$ becomes gaussian.  It can thus be performed exactly to yield
\begin{align}
\begin{split}
    Z_{n} &= 8 \pi \int d\mathcal{R} \ \mathcal{R} e^{\pi \mathcal{R}^2 - \beta \left(
    \frac{\mathcal{R}^3}{2L^2} + \frac{\mathcal{R}}{2}
    \right)} \int dQ e^{- \beta\left(
\frac{Q^2}{2\mathcal{R}} - \mu_n Q
    \right)}\\ &= 8 \sqrt{2} \pi^{3/2} \beta^{-1/2} \int d\mathcal{R} \ \mathcal{R}^{3/2} \exp \left(
- \beta \left(
\frac{\mathcal{R}^3}{2L^2} + \frac{\mathcal{R}}{2} \left(
1-\mu_n^2
\right)
\right) + \pi \mathcal{R}^2
    \right).
\end{split}
\label{Zn0}
\end{align}
Since the exponent on the final line is a cubic polynomial
\begin{equation}
\label{eq:uR}
    u(\mathcal{R}) = - \beta \left(
\frac{\mathcal{R}^3}{2L^2} + \frac{\mathcal{R}}{2} \left(
1-\mu_n^2
\right)
\right) + \pi \mathcal{R}^2,
\end{equation}
one might naturally think that \eqref{Zn0} would have asymptotic behavior similar to that of an Airy function (for which the contour of integration in \eqref{Zn0} would run over the full imaginary axis). However, that conclusion would be premature because the integral over $\mathcal{R}$ in fact runs only over the positive half-line. This is not just a mere technicality.  Since the argument of the exponent is a cubic polynomial with real leading coefficient, integrating over negative values of $\mathcal{R}$ would have given a divergent answer. In addition, we see that the integral has a non-analyticity at $\mathcal{R} = 0$. As a result, Cauchy's theorem allows only deformations of the integration contour that preserve the endpoint at $\mathcal{R} = 0$. 

Before analyzing \eqref{Zn0} in detail, it is useful to consider what sort of contributions we expect to find in the semiclassical limit.   In particular, in addition to saddle point contributions, we may also find endpoint contributions from $\mathcal{R} = 0$.  (There can be no endpoint contribution from $\mathcal{R} = \infty$ since  the integrand vanishes in the limit  $\mathcal{R} \rightarrow +\infty$.) Note that, when $\Re \mu_n^2 <1$, the magnitude of our integrand is larger at the endpoint ${\mathcal R}=0$ than at other nearby points on the positive real line.  In this case, endpoint contribution at $\mathcal{R}=0$ can be approximated by 
\begin{equation}
   8 \sqrt{2} \pi^{3/2} \beta^{-1/2} \int_{\mathbb{R}_{>0}} d\mathcal{R} \ \mathcal{R}^{3/2} \exp\left(u(0) + u'(0) \mathcal{R} \right) = \frac{48 \pi^2}{\beta^3 (1-\mu_n^2)^{5/2}}.
   \label{ref:boundary_contr}
\end{equation}
In particular, in this case the above integral converges.  Here our \eqref{ref:boundary_contr} gives a precise numerical coefficient and so includes information at the level of what would be called a one-loop correction around a saddle.

It turns out that \eqref{ref:boundary_contr} in fact gives a useful notion of an endpoint contribution even  when 
$\Re \mu_n^2 <1$.  This can be seen by using a version of steepest descent flow (suitably modified to leave fixed the endpoint at $\mathcal {R}=0$) to deform the integration contour to a set of Lefschetz thimbles.  One expects the resulting final contour to consist of steepest descent contours through the relevant saddles as well as an additional contour following the steepest descent path from the endpoint $\mathcal{R}=0$.  Performing the analogue of the integral in \eqref{ref:boundary_contr} over this latter contour gives the right hand side of 
\eqref{ref:boundary_contr}  for all values of $\mu_n$.  Here one should note that, since ${\mathcal R}^{3/2}$ has a branch point at the endpoint ${\mathcal R=0}$, these functions live on a 2-sheeted Riemann surface.  There are thus in fact two steepest descent paths from ${\mathcal R=0}$, with one on each sheet, and with the relevant such path being determined by the above-mentioned steepest descent flow. 

Let us now study the (possible) saddle point contributions.
A cubic exponent typically leads to two saddles, and these saddles are generally complex:
\begin{subequations}
\begin{equation}
    \mathcal{R}_\pm = \frac{L \left(2\pi L \pm \sqrt{3 \beta ^2 \left(\mu_n ^2-1\right)+4 \pi ^2 L^2}\right)}{3 \beta },  
    \end{equation}
    with
    \begin{equation}
    u(\mathcal{R}_\pm) = \frac{L \left(\sqrt{3 \beta ^2 \left(\mu_n ^2-1\right)+4 \pi ^2 L^2} \pm 2 \pi  L\right) \left(\pi  L \sqrt{3 \beta ^2 \left(\mu_n^2-1\right)+4 \pi ^2 L^2}\pm2 \pi ^2 L^2 \pm 3 \beta ^2 \left(\mu_n ^2-1\right)\right)}{27 \beta ^2}.
    \end{equation}
    \label{ref:saddles}
\end{subequations}
For real $\mu_n$ (and thus, for us, for $n=0$) with $4\pi^2 L^2\ge 3\beta^2(1-\mu_n^2)$, the choice of sign $\pm$ selects either the large or small black hole; i.e., ${\cal R}_\pm$ are both real and we have ${\mathcal R}_+ \ge {\mathcal R}_-$ (though ${\mathcal R}_-$ is negative for $\mu_n^2<1$). We therefore expect the ${\mathcal R}_+$ saddle to be relevant and the ${\mathcal R}_-$ saddle to be irrelevant.  Indeed, this is clear from the fact that, in this case, for ${\mathcal R} \in {\mathbb R}^+$ the exponent $u$ is maximized\footnote{In particular, we see from \eqref{ref:saddles} that $u({\mathcal R}_+)$ is non-negative when ${\mathcal R}_+$ is real, so that $u({\mathcal R}_+)\ge u(0)=0$.} at  ${\mathcal R}_+$, while $u$ has a local minimum at ${\mathcal R}_-$ (which also fails to lie on the integration contour for $\mu_n^2<1$).  However, the situation is much less clear for complex $\mu_n$ since then both ${\cal R}_\pm$ become complex and, in fact, one can exchange ${\cal R}_\pm$ by following a closed loop in the complex $\mu_n$-plane which wraps around the branch point.  There can thus be no sharp distinction between the large and small black hole branches for general complex $\mu_n$.

In seeking to better understand the relevance of complex saddles, it turns out to be  instructive to examine the asymptotic behavior at large $n$.  Taking (recall, real) $\mu_0>0$ and $n \to \pm \infty$ yields
\begin{subequations}
    \begin{equation}
        \mathcal{R} = \pm \frac{2 i \pi  L n}{\sqrt{3} q \beta }+\frac{L \left(\pm \sqrt{3} \beta  \text{$\mu_0$}+ 2 \pi  L\right)}{3 \beta }+O\left(\frac{1}{n}\right)
    \end{equation}
    \begin{equation}
        u =\mp \frac{8 i \pi ^3 L n^3}{3 \sqrt{3} q^3 \beta ^2} -\frac{4 n^2 \left(\pi ^2 L \left( \pm\sqrt{3} \beta  \text{$\mu_0$}+\pi  L\right)\right)}{3 q^2 \beta ^2} + O\left(n\right).
    \end{equation}
\end{subequations}
Let us recall again that, in a Picard-Lefschetz analysis, a saddle contributes if and only if the steepest {\it ascent} contour from the saddle has non-zero intersection number with the contour of integration.  Steepest ascent contours are, in particular, contours of constant phase.  It is therefore useful to note that the phase of our integrand at each saddle is determined by the imaginary part of $u$, and that at large $|n|$, we have $\textrm{sgn} \Im u = \mp \textrm{sgn} \ n$. 

On the other hand, even at finite $n$, from \eqref{eq:uR} the phase of the integrand along the integration contour is $\rm{Im}\, u = 2 \pi n q^{-1} \mathcal{R} \mu$. Since $\mathcal{R} \in (0, \infty)$ and we took $\mu>0$, this phase has the same sign as $n$. It follows that the ascent curve from the ${\mathcal R}_+$ saddle cannot cross the contour of integration and that this saddle (which one might have naively called the `large black hole saddle') cannot contribute in the limit of large $n$.

Let us now look more closely at the ${\mathcal R}_-$ saddle. Note that at sufficiently low temperatures (namely, for $\pi L - \sqrt{3} \beta \mu_0<0$) and large $|n|$, it would provide a contribution that is exponentially large in $n$ (as found previously in \cite{Harlow2022}).   In particular, it would give  $|Z_{n}|>Z_{0}$ which we have already seen to be forbidden. In a moment we will see in detail why the integration contour cannot be usefully deformed to pass through such saddles. For now, we simply remind the reader that, in the saddle-point approximation and for a relevant saddle, $\Re u$ at the saddle cannot exceed the maximum of $\Re u$ along the integration contour, and that this would clearly be violated if the ${\mathcal R}_-$ saddle were to contribute for large positive $n$ and $\mu>0$.  

We thus see that in the large $|n|$ limit\footnote{Which here means $|\frac{n}{q}|\gg \beta \sqrt{\mu_0^2+1}$ and $|\frac{n}{q}|\gg L$ in units of the Planck length. 
}, and for either sign of $n$, the integral is dominated by the endpoint contribution.  In particular, we then have $\Re \mu_n^2$ negative, and thus $\Re \mu_n^2 < 1$. From  \eqref{ref:boundary_contr} we find
\begin{equation}
    Z_{n} \approx \frac{48 \pi^2}{\beta^3 (1-\mu_n^2)^{5/2}}.
\end{equation}
The final partition function is a sum over all $n$ and, since $Z_{n}$ decays as $n^{-10}$, the sum is clearly convergent, at least in the approximation used in this work\footnote{Here we have included what would be called 1-loop effects association with the integrals in \eqref{eq:Znm}, though we have ignored other one-loop effects.  We do not mean to imply that such an approximation is justified, or that it can be trusted.  We simply hope that the one-loop effects so included may indicate qualitative features of other 1-loop effects that would be seen in a more complete analysis.}. This then resolves the puzzle raised in \cite{Harlow2022}. 

Given the above comments, it may be natural to wonder whether all of the complex saddles may be forbidden from being relevant.  But it is easy to see that this cannot be the case.  In particular, let us recall that for $\mu_0 \ge 1$ the function $u$ is real on the integration contour and is maximized on this contour at  ${\mathcal R} = {\mathcal R}_+$ for all $\beta >0 $.  As a result, the 
${\mathcal R}_+$ saddle clearly contributes (and in fact dominates) in this case.  This means that the ascent contour through ${\mathcal R_+}$ has non-zero intersection number with the integration contour.  But such intersection numbers are topological, and are thus stable under small deformations.\footnote{The only exception is the occurance of Stokes' phenomena, which requires there to be a second intervening saddle on the ascent contour between the saddle of interest and the contour of integration.  But this cannot occur here since the ${\mathcal R}_+$ saddle lies on the integration contour, leaving no room for a distinct saddle to intervene.}  Note then that since $\mu_n$ differs from $\mu_0$ only by $\frac{2\pi n}{q\beta} i$, we have $\mu_n \rightarrow \mu_0$ at fixed $n$ in the limit of large $\beta$.  Taking $\beta$ large but finite with $n,L$ fixed, we must thus find the contours for $\mu_n$ to be small deformations of those for $\mu_0$ so that the intersection numbers agree and the ${\mathcal R}_+$ saddle remains relevant.  Indeed, by continuity it must also remain dominant over all other contributions to $Z_n$ at sufficiently large values of $\beta$ (though we will still have $|Z_n|<Z_0$ by \eqref{eq:Z00}). 

 Additionally, the analytic analysis in appendix \ref{app:analytic} shows that at small $\beta$ (and arbitrary $\mu_0$), all ${\mathcal R}_+$  saddles  with $Lq > \sqrt{3}n$ contribute to our partition function. One should expect this to be the case since, in this regime $\Re \mathcal{R}_+$ is very large (of order $\beta^{-1}$) while $\Im \mathcal{R}_+$ is of order $\beta^0$.  We may therefore expect that such saddles behave much like the familiar large black hole (${\mathcal R}_+$) saddle with $n=0$.  As a result, at large temperatures, there is a finite (possibly zero) number of non-perturbative, though still exponentially large, corrections to the leading semiclassical partition function that are associated with these additional saddles. It would be interesting to see if  similar effects occur in top-down string-theoretic models associated with known gauge-theory duals, and if they can be reproduced from the dual gauge  theory.


In summary, for  large enough
$n$
(and, in particular for $n \gg \beta q$ when $\beta \gg L^2$ and $\mu_0>1$) there can be no contributions from saddle points, though such contributions must arise for  $\beta \mu_0 \gg n$ with $\beta \gg L^2$ and $\mu_0>1$. 
Analytic arguments that saddles (with $\mu_0>1$ do indeed begin to contribute
for $\beta q$ of order $n$ are given in Appendix \ref{app:analytic}.
Since it is instructive to see more precisely how the relevance of a given saddle changes with $\beta$, we also study this question numerically for the case $\mu_0 >1$ in section \ref{sec:CP}, and more briefly for the case $\mu_0<1$ in section \ref{sec:smallmu}.

\subsection{$\mu_0>1 $}
\label{sec:CP}
We will now describe the  relevant contours in detail. Since $u$ is a cubic polynomial, constant phase curves are given by cubic planar curves. As such, it is rather cumbersome to treat the problem analytically. Instead, we will simply plot the relevant quantities for a few representative values of the parameters. Analytic expressions for the ascent and decent contours are given in the Appendix.

\begin{figure}[htbp]
    \centering
    \begin{subfigure}[b]{0.45\textwidth}
        \centering
        \includegraphics[width=\linewidth]{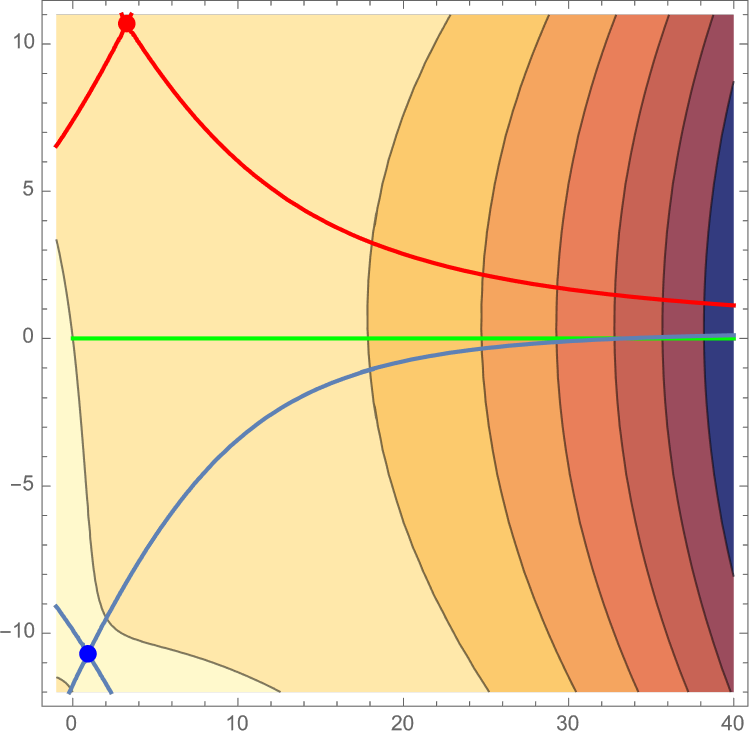}
        \caption{$\beta L = 1$}
    \end{subfigure}
    \hfill
    \begin{subfigure}[b]{0.45\textwidth}
        \centering
        \includegraphics[width=\linewidth]{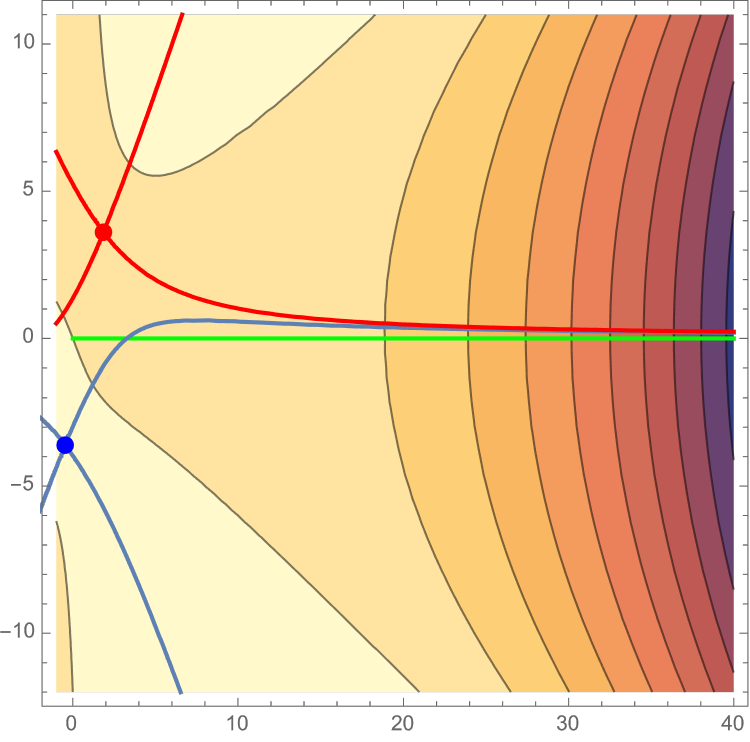}
        \caption{$\beta L = 3$}
    \end{subfigure}
    \newline
    \centering
    \begin{subfigure}[b]{0.45\textwidth}
        \centering
        \includegraphics[width=\linewidth]{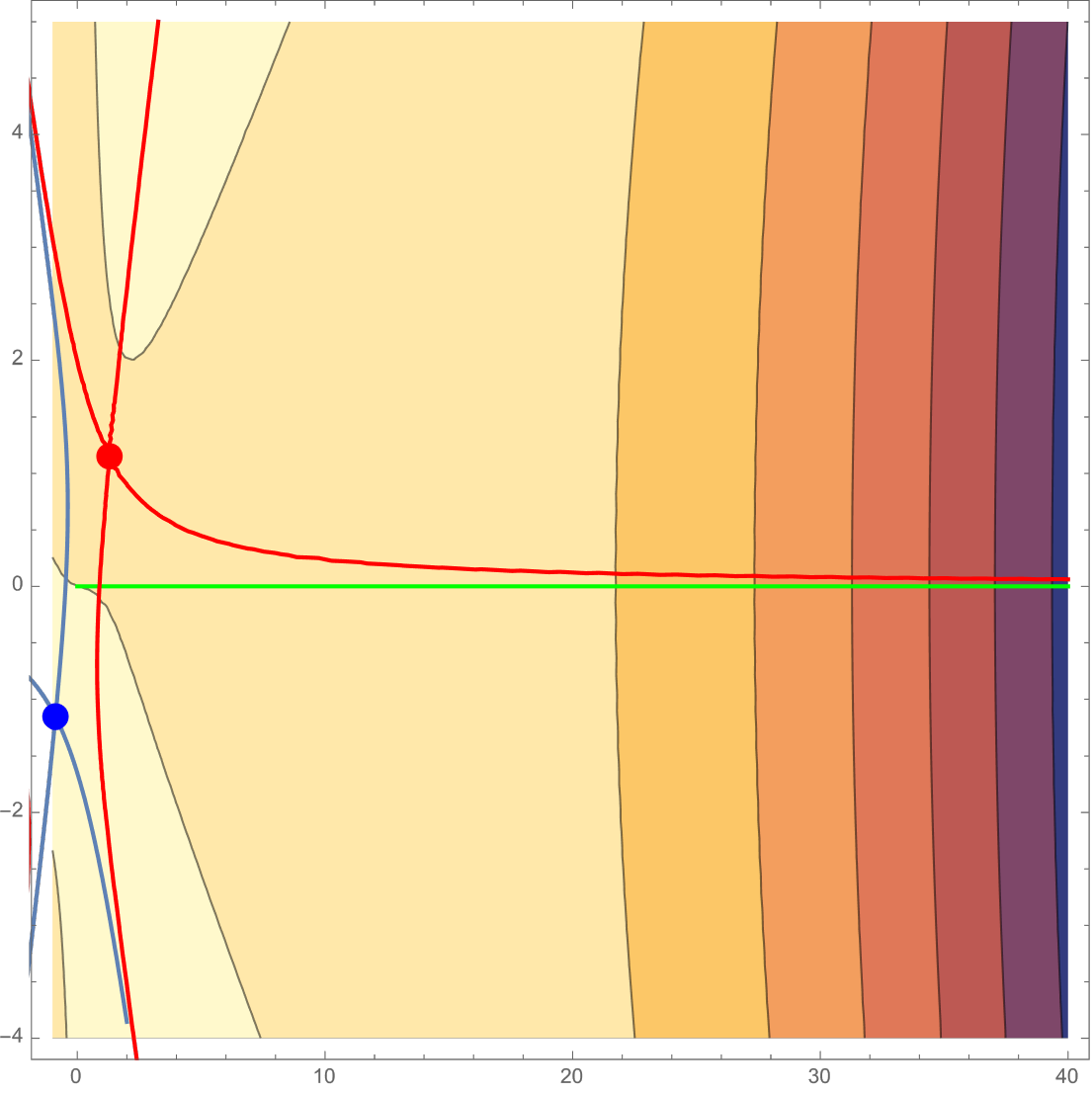}
        \caption{$\beta L = 10$}
        \label{fig:mu2_both_contr}
    \end{subfigure}
    \hfill
    \begin{subfigure}[b]{0.45\textwidth}
        \centering
        \includegraphics[width=\linewidth]{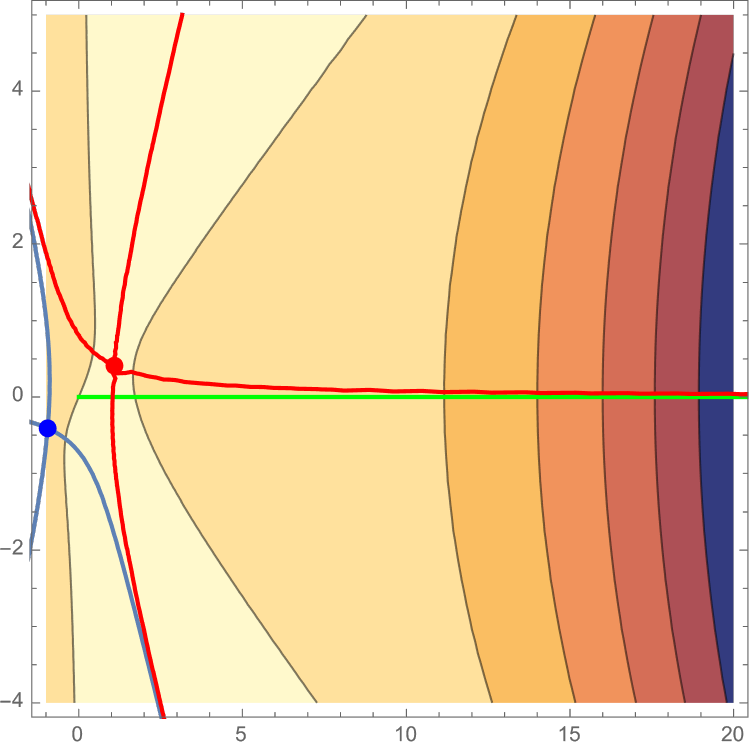}
        \caption{$\beta L = 30$}
    \end{subfigure}
    \caption{The real part of $u$ for $\mu_0 = 2$ and $\frac{n}{qL}=3$, with darker and redder shading indicating more negative values of $\Re \, u$. The panels a,b,c,d show a sequence of plots with increasing $\beta L$.  The green line represents the contour of integration $r_+ \in (0,\infty)$. The blue and red curves are constant phase curves for the  corresponding saddles (blue and red dots, with blue denoting ${\mathcal R}_-$ and red denoting ${\mathcal R}_+$). The red steepest ascent line crosses the green contour in the final two panels, but not in the first two.   This is consistent with the expectation that, for $qL = \frac{n}{3}< \sqrt{3}n$  and $\mu_0 \ge 1$, the red ($\mathcal{R}_+$) saddle should contribute at large $\beta$ but not at small $\beta$. In agreement with out analytic predictions, we also find the red saddle to approach the real line as $\beta \to \infty$. The blue ($\mathcal{R}_-$) saddle never contributes since the shading indicates that the blue line that crosses the real axis is the blue  steepest {\it descent} line.}
    \label{fig:mu2}
\end{figure}

For concreteness, we will take $\mu_0 = 2$, $n=3$, $L=1$ and $q=1$ and study how the desired contours change as a function of temperature. Fig. \ref{fig:mu2} shows the saddles and associated steepest ascent/descent contours across various temperatures. We find that at small $\beta$ (large temperature) this saddle does not contribute, start contributing when $\beta q \sim n$ and then continues up to arbitrarily large $\beta$ (small temperature). As predicted above, we also find numerically that this saddle approaches the real axis  as $\beta$ grows. The \ref{fig:mu2_both_contr} panel shows an interesting case in which both the saddle and the endpoint contribute (this case satisfies $\Re \mu_n^2 <1$). In particular, the latter contribution is orders of magnitude larger than the former and so cannot be neglected.

\begin{figure}[htbp]
    \centering
    \begin{subfigure}[b]{0.45\textwidth}
        \centering
        \includegraphics[width=\linewidth]{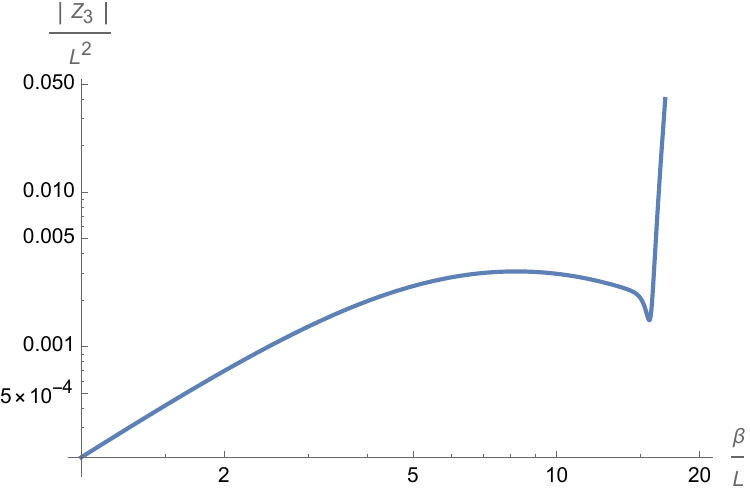}
        \caption{Numerically integrated $Z_3$}
    \end{subfigure}
    \hfill
    \begin{subfigure}[b]{0.45\textwidth}
        \centering
        \includegraphics[width=\linewidth]{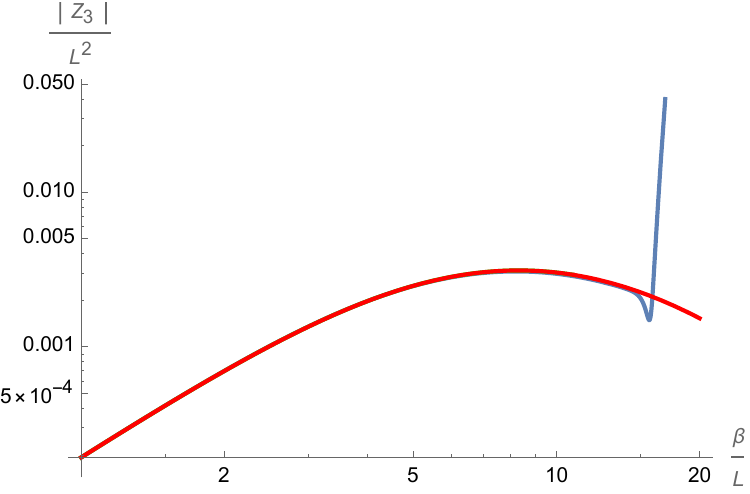}
        \caption{Numerically integrated $Z_3$ and the boundary contribution}
    \end{subfigure}
    \newline
    \centering
    \begin{subfigure}[b]{0.45\textwidth}
        \centering
        \includegraphics[width=\linewidth]{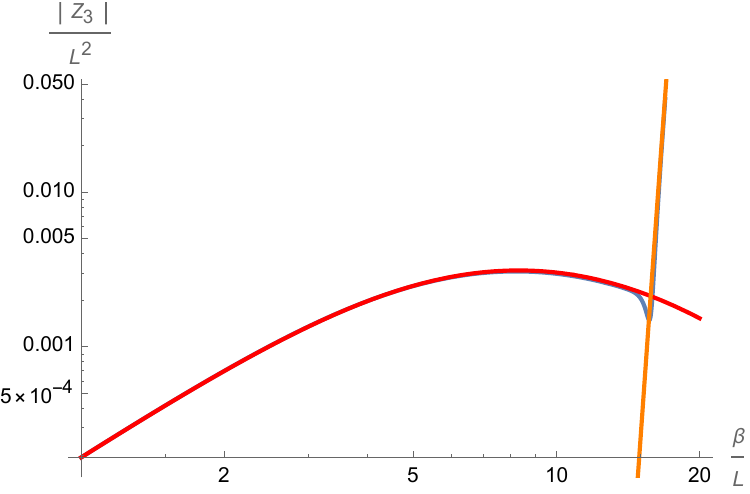}
        \caption{Numerically integrated $Z_3$, the boundary contribution and the saddle contribution}
    \end{subfigure}
    \hfill
    \begin{subfigure}[b]{0.45\textwidth}
        \centering
        \includegraphics[width=\linewidth]{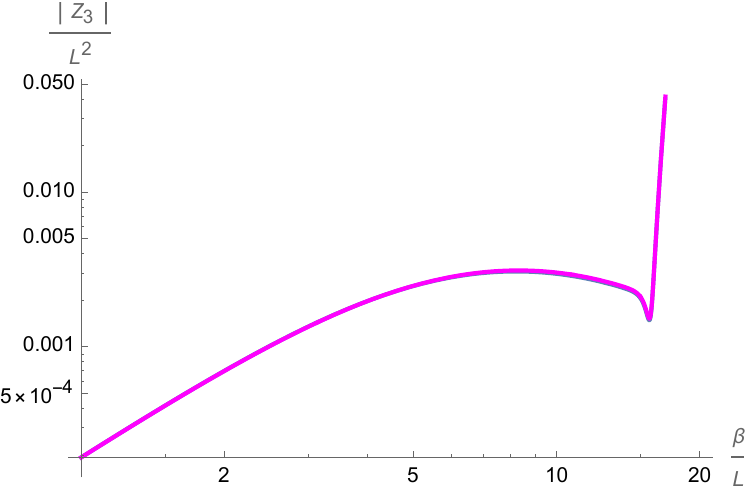}
        \caption{Numerically integrated $Z_3$ and the boundary plus saddle contributions}
    \end{subfigure}
    \caption{The panels present $Z_3$ evaluated at $\hbar = 1, \mu_0 =2, n=3$ and $qL=1$. We believe that the sudden dip around $\beta \sim 17L$ is likely to be a result of working at finite $\hbar$, and that it would be replaced by a sharp phase transition in the semiclassical limit. In the left-top panel, we present the absolute value (blue) of the result of numerically integrating \eqref{Zn0}. The right-top panel overplots the absolute value (red) of the boundary contribution \eqref{ref:boundary_contr}. The left-bottom panel further overplots the absolute value (orange) of the contribution from the $\mathcal{R}_+$ saddle. Finely, the right-bottom panel sums the boundary contribution and the ${\mathcal R}_+$ saddle including one-loop corrections for both.  Despite the fact that we work at $\hbar=1$, the result (magenta) completely covers the blue numerical curve, rendering the latter invisible.  In particular,  reproducing the dip required the correct phase difference between these two contributions.  It thus serves as a check at the one-loop level.}
    \label{fig:numerical_int}
\end{figure}

To illustrate the relative contributions of saddles and endpoints, we have also computed the partition function by direct numerical integration for several sets of parameters.  The results are shown in Fig. \ref{fig:numerical_int}. In the right-top panel we compare this with the endpoint contribution \eqref{ref:boundary_contr} alone, while  tne bottom panels also add a one-loop-corrected saddle-point contribution to \eqref{eq:Znm}. As one can see, even though we work at $\hbar =1$, the match is surprisingly good. In particular, there is a region where these two actions are comparable and one need to sum them to obtain good agreement with numerics. In general we see that both saddle and endpoint contributions can be important and that neither can be generally neglected.

Although we presented plots only for $\mu_0 =2$ and $n=3$, similar conclusions hold for all $\mu_0^2 > 1$ and for large $n \neq 0$. However, we find different behavior for saddles that satisfy $L q> \sqrt{3}|n|$. For these, the $\mathcal{R}_+$  saddles contribute at all temperatures and there is no Stokes phenomenon. Nevertheless, if $2|n|> Lq > \sqrt{3}|n|$, 
at large temperatures (small $\beta$) the saddle contribution is exponentially surpassed relative to the endpoint, while the saddle dominates at large temperatures (small $\beta$).   As a result, the phase transition seen at large $n$ still occurs in this regime.  For $Lq >2|n|$ the saddle instead dominates at all temperatures.  These results follow from the analytic formulae in appendix \ref{app:analytic}.

\subsection{$\mu_0 \le 1$}
\label{sec:smallmu}
We now describe the contributions of  saddles and endpoints for the case $\mu_0\le 1$.

 \begin{figure}[htbp]
    \centering
    \begin{subfigure}[b]{0.45\textwidth}
        \centering
        \includegraphics[width=\linewidth]{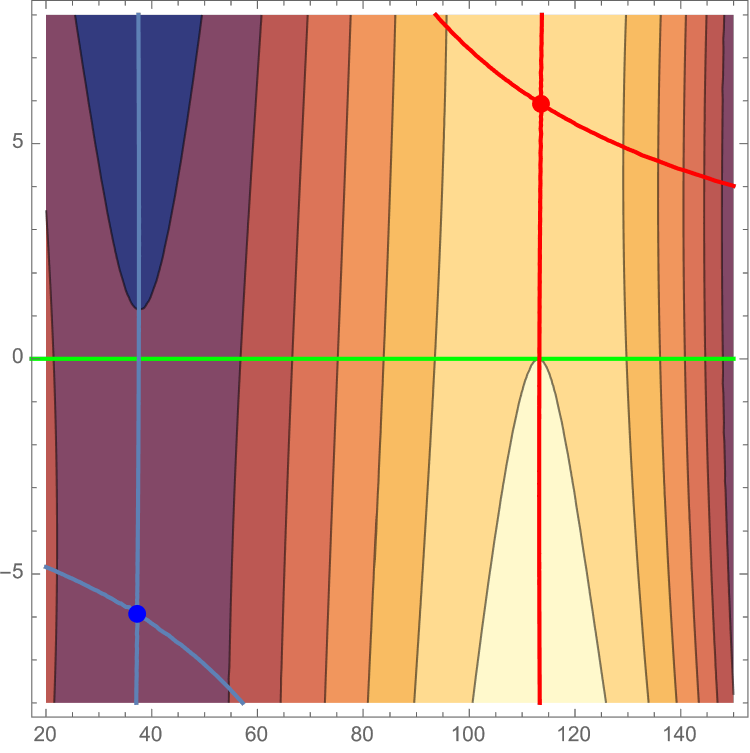}
        \caption{$\beta L = \frac{1}{6}$}
    \end{subfigure}
    \hfill
    \begin{subfigure}[b]{0.45\textwidth}
        \centering
        \includegraphics[width=\linewidth]{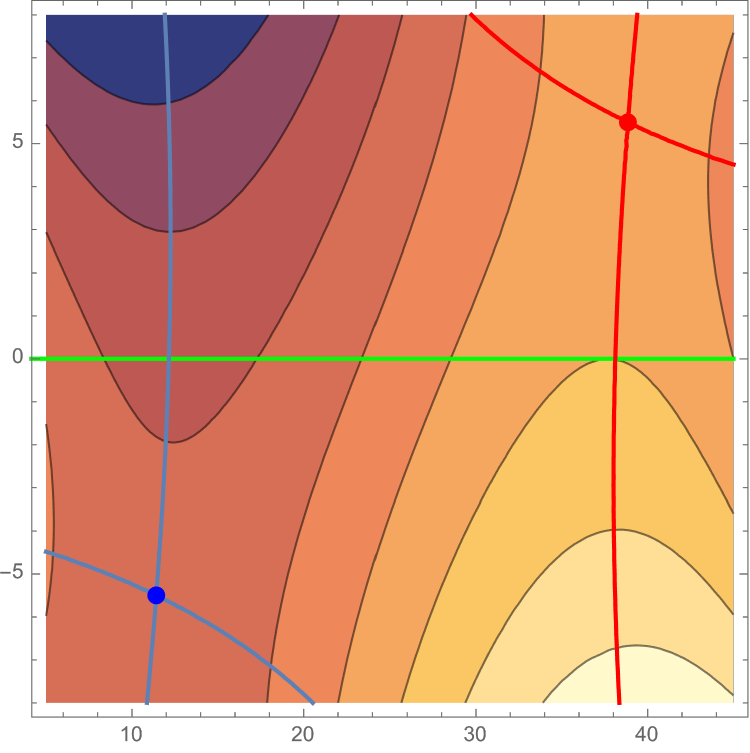}
        \caption{$\beta L = \frac{1}{2}$}
    \end{subfigure}
    \newline
    \centering
    \begin{subfigure}[b]{0.45\textwidth}
        \centering
        \includegraphics[width=\linewidth]{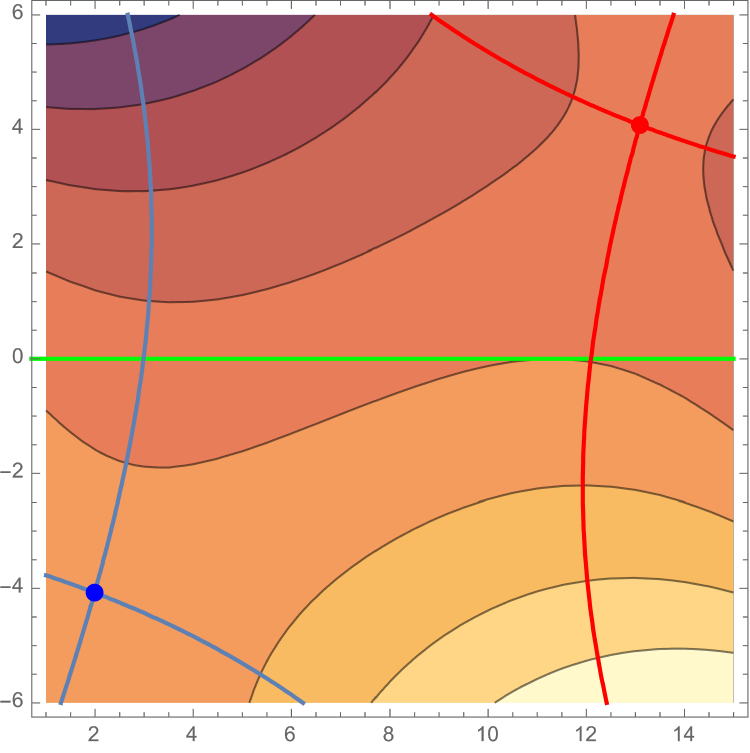}
        \caption{$\beta L = \frac{5}{3}$}
    \end{subfigure}
    \hfill
    \begin{subfigure}[b]{0.45\textwidth}
        \centering
        \includegraphics[width=\linewidth]{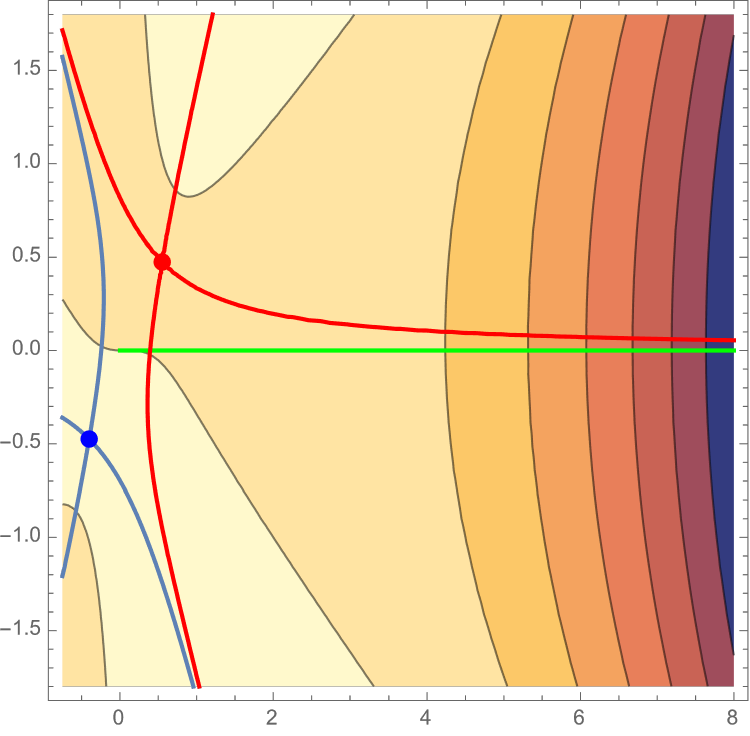}
        \caption{$\beta L = \frac{500}{3}$}
    \end{subfigure}
    \caption{The panels show the real part of $u$ for $\mu_0 = 1$ and $\frac{n}{qL}=\frac{1}{2}$ using the same color coding as in Fig. \ref{fig:mu2}. The sequence $a,b,c,d$ again shows increasing $\beta L$.  Since the red steepest ascent line crosses the green contour in each panel, the red (${\mathcal R}_+$) saddle always contributes. This is consistent with the expectation that for $qL = 2n> \sqrt{3}n$ and $\mu_0 \ge 1$, the red ($\mathcal{R}_+$) saddle should contribute at both small and large $\beta$. The blue ($\mathcal{R}_-$) saddle never contributes since the shading indicates that only its steepest {\it descent} line ever crosses the green contour.}
    \label{fig:mu1_n3}
\end{figure}

 \begin{figure}[htbp]
    \centering
    \begin{subfigure}[b]{0.45\textwidth}
        \centering
        \includegraphics[width=\linewidth]{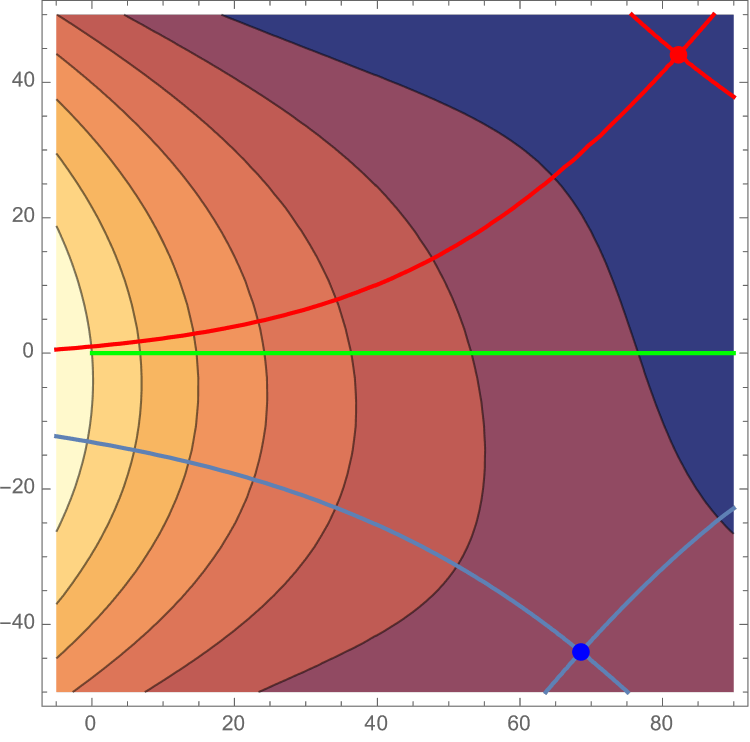}
        \caption{$\beta L = \frac{1}{6}$}
    \end{subfigure}
    \hfill
    \begin{subfigure}[b]{0.45\textwidth}
        \centering
        \includegraphics[width=\linewidth]{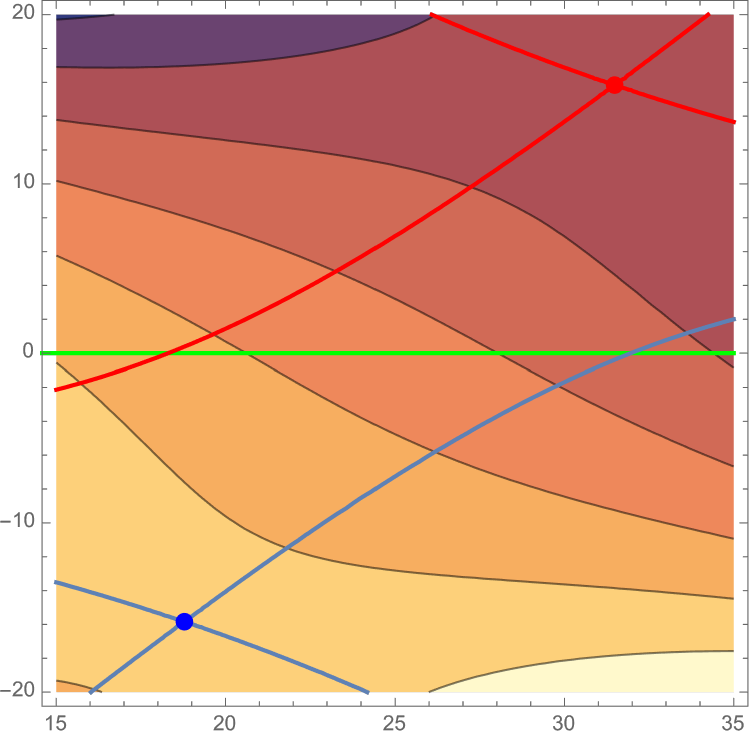}
        \caption{$\beta L = \frac{1}{2}$}
    \end{subfigure}
    \newline
    \centering
    \begin{subfigure}[b]{0.45\textwidth}
        \centering
        \includegraphics[width=\linewidth]{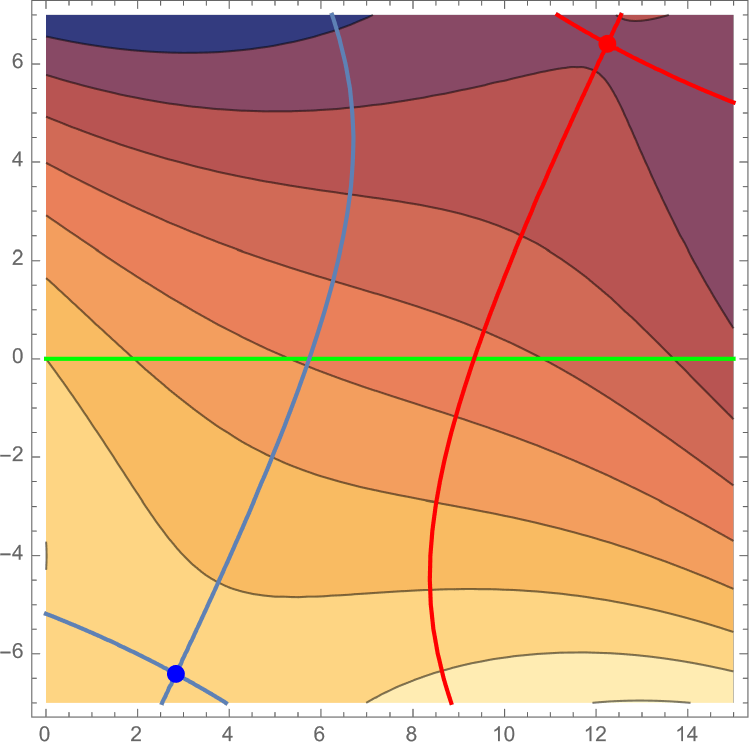}
        \caption{$\beta L = \frac{5}{3}$}
    \end{subfigure}
    \hfill
    \begin{subfigure}[b]{0.45\textwidth}
        \centering
        \includegraphics[width=\linewidth]{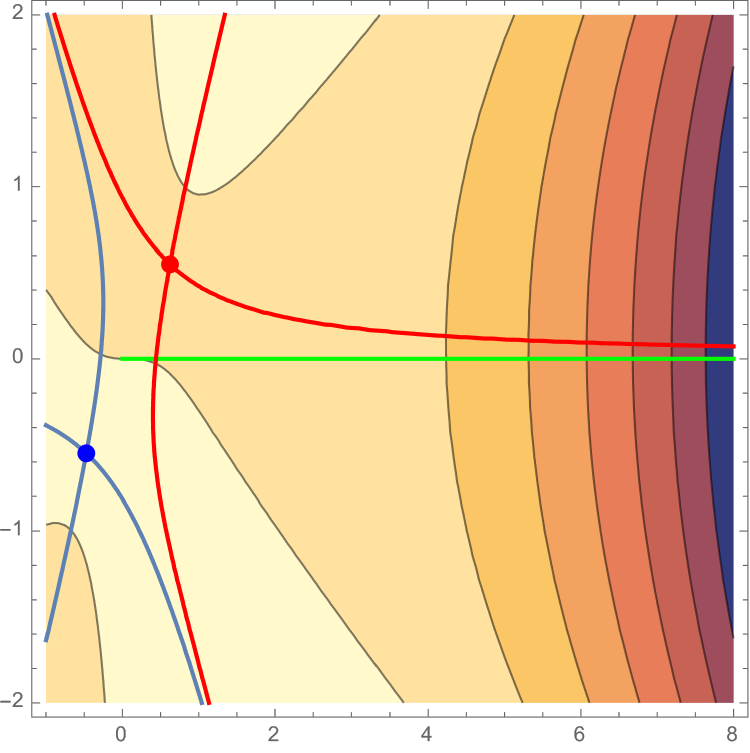}
        \caption{$\beta L = \frac{500}{3}$}
    \end{subfigure}
    \caption{The panels show the real part of $u$ for $\mu_0 = 1$ and $\frac{n}{qL}=\frac{2}{3}$ using the same color coding as in Fig. \ref{fig:mu2}.  The sequence a,b,c,d shows increasing $\beta L$. The red steepest ascent line crosses the green contour in the last three panels, though not in the first panel.   This is consistent with the expectation that, for $qL = \frac{3}{2}n< \sqrt{3}n$ and $\mu_0=1$, the red ($\mathcal{R}_+$) saddle will fail to contribute at small $\beta$ but should instead contribute at larger $\beta$. The blue ($\mathcal{R}_-$) saddle never contributes since only its steepes {\it descent} line ever crosses the green contour.}
    \label{fig:mu1_n4}
\end{figure}

Let us begin with the borderline case  
$\mu_0=1$ in which the (real) black hole solution exists for all $\beta$ but does not approach a large real Euclidean black hole as $\beta \to \infty$. Instead, we have  $\mathcal{R}_+ = 
\frac{4\pi L^2}{3\beta} \to 0$). As a result, we cannot repeat our previous argument that more and more saddles should contribute in the limit of large $\beta$. Nevertheless, we find the conclusion to remain true, as seen in Fig. \ref{fig:mu1_n3}, \ref{fig:mu1_n4}. This can be traced to the fact that this limit yields $\Re \mathcal{R}_+ \sim \Im \mathcal{R}_+ \sim \beta^{-1/2}$ and thus that the saddles remain close to the contour of integration.

For $\mu_0 <1$ we find a plethora of different behaviors. Let us summarize them quickly:
\begin{itemize}
    \item At large temperatures ($\beta \to 0$), we again find behavior analogous to that for $\mu_0>1$. There is a finite number of ${\mathcal R}_+$ saddles that contribute in this limit, and they are again determined by the condition $L q > \sqrt{3} |n|$. The saddle dominates the calculation of $Z_n$ if $Lq> 2 |n|$. Otherwise, it is subdominant with respect to the boundary contribution. 
    \item However, at low temperatures ($\beta \to \infty$), a given saddle {\it never} contributes.  Instead, in this limit $Z_n$ is well-approximated by the boundary contribution for any $n$.
    \item The behavior of saddles is no longer monotonic with $\beta$.  By this we mean that there are saddles that contribute in some finite interval, though they fail to contribute at both small and large $\beta$. However, such saddles never dominate.\footnote{The magnitude of the integrand turns out to be greatest at the ${\mathcal R}=0$ endpoint.  Saddles for which their ascent contour intersects the integration contour must thus also have magnitude smaller than that at the endpoint.}
\end{itemize}

    \begin{figure}[htbp]
    \centering
    \begin{subfigure}[b]{0.35\textwidth}
        \centering
        \includegraphics[width=\linewidth]{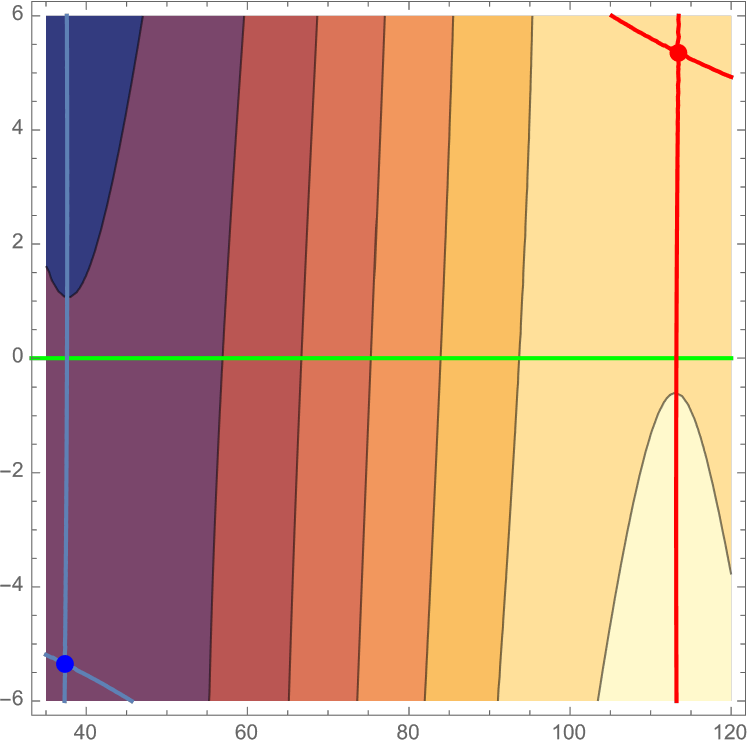}
        \caption{$\beta L = \frac{1}{6}$}
    \end{subfigure}
    \hfill
    \begin{subfigure}[b]{0.35\textwidth}
        \centering
        \includegraphics[width=\linewidth]{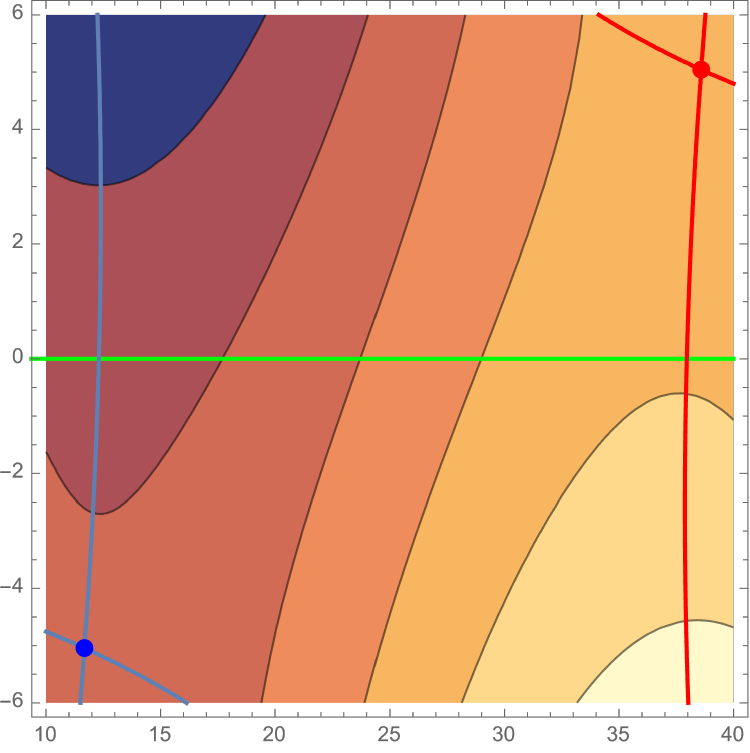}
        \caption{$\beta L = \frac{1}{2}$}
    \end{subfigure}
    \newline
    \centering
    \begin{subfigure}[b]{0.35\textwidth}
        \centering
        \includegraphics[width=\linewidth]{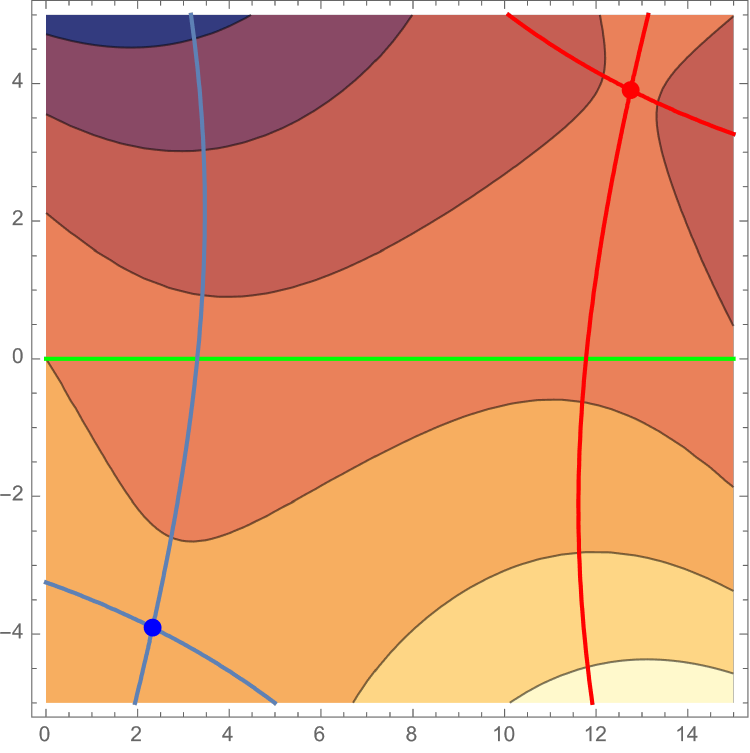}
        \caption{$\beta L = \frac{5}{3}$}
    \end{subfigure}
    \hfill
    \begin{subfigure}[b]{0.35\textwidth}
        \centering
        \includegraphics[width=\linewidth]{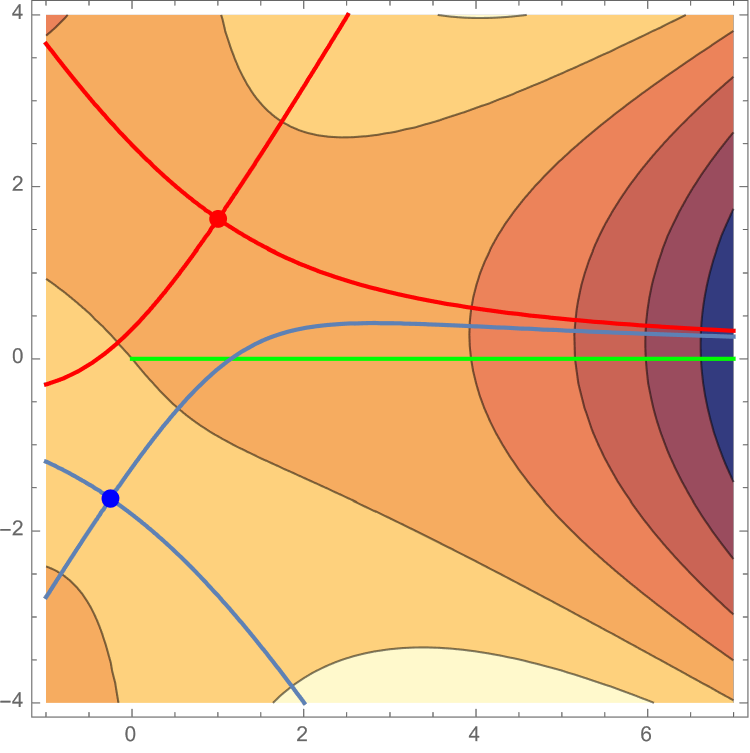}
        \caption{$\beta L = \frac{100}{3}$}
    \end{subfigure}
    \caption{The panels show the real part of $u$ for $\mu_0 = 0.9$ and $\frac{n}{qL}=\frac{1}{2}$ using the same color coding as in Fig. \ref{fig:mu2}. In the first three panels the red steepest ascent line crosses the green contour, though this does not happen in the  last panel.  This  illustrates the fact that for $\mu_0<1$ the ${\mathcal R}_+$ saddle fails to contribute at small temperatures.   In particular, it is consistent with the expectation that, for $qL = 2n> \sqrt{3}n$ and $\mu_0< 1$, the red ($\mathcal{R}_+$) saddle contributes at small $\beta$ but cannot contribute at large $\beta$. The blue ($\mathcal{R}_-$) saddle never contributes since only its steepest descent line ever  crosses the green contour.}
    \label{fig:mu09_n3}
\end{figure}

    \begin{figure}[htbp]
    \centering
    \begin{subfigure}[b]{0.45\textwidth}
        \centering
        \includegraphics[width=\linewidth]{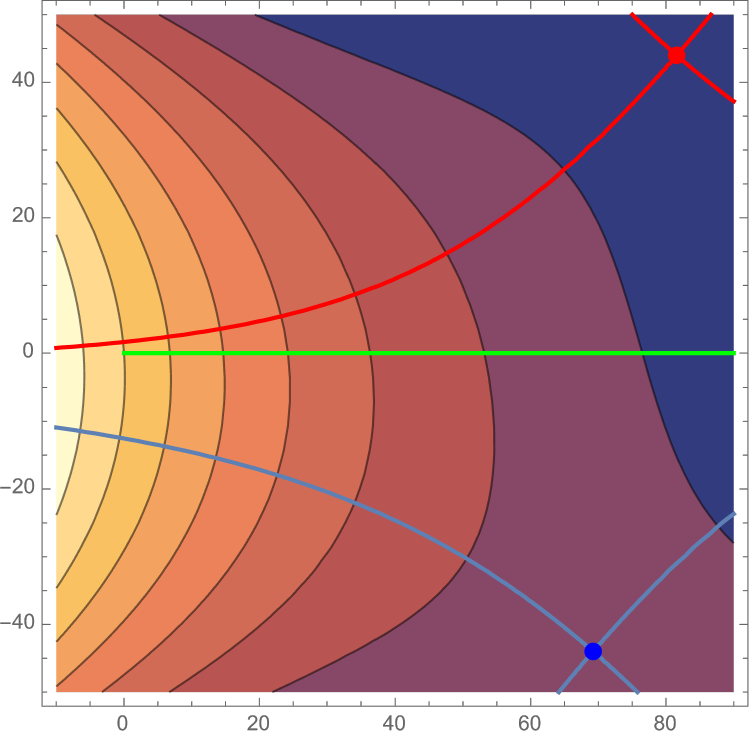}
        \caption{$\beta L = \frac{1}{6}$}
    \end{subfigure}
    \hfill
    \begin{subfigure}[b]{0.45\textwidth}
        \centering
        \includegraphics[width=\linewidth]{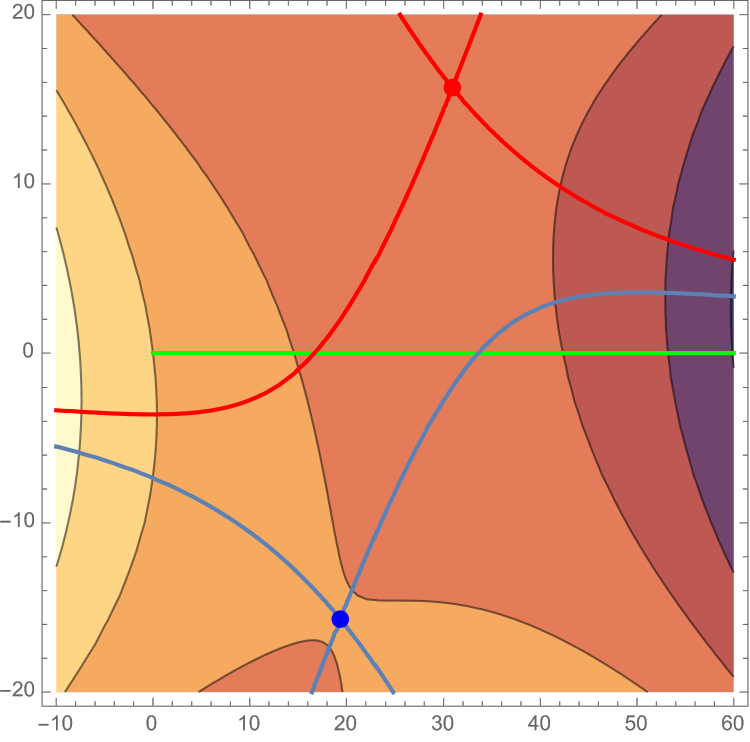}
        \caption{$\beta L = \frac{1}{2}$}
    \end{subfigure}
    \newline
    \centering
    \begin{subfigure}[b]{0.45\textwidth}
        \centering
        \includegraphics[width=\linewidth]{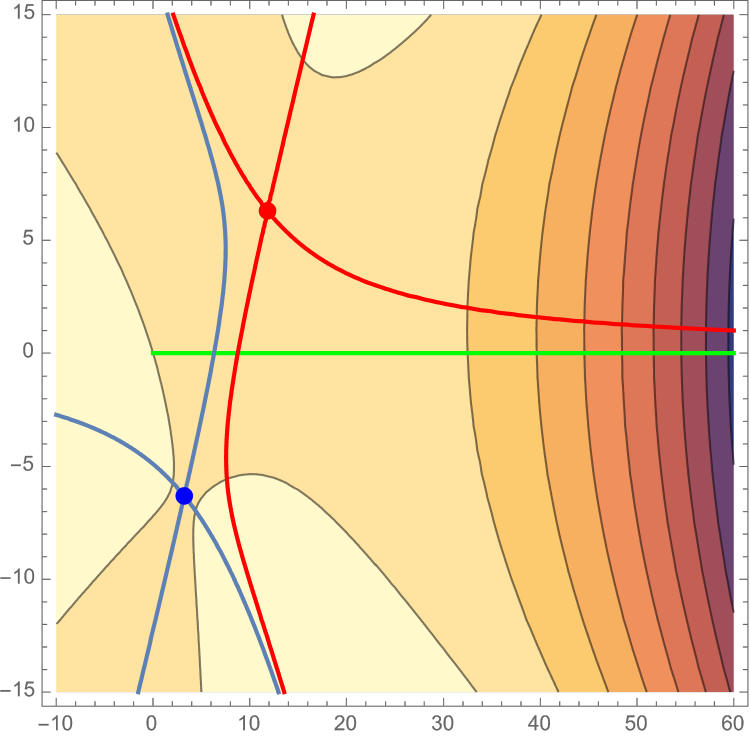}
        \caption{$\beta L = \frac{5}{3}$}
    \end{subfigure}
    \hfill
    \begin{subfigure}[b]{0.45\textwidth}
        \centering
        \includegraphics[width=\linewidth]{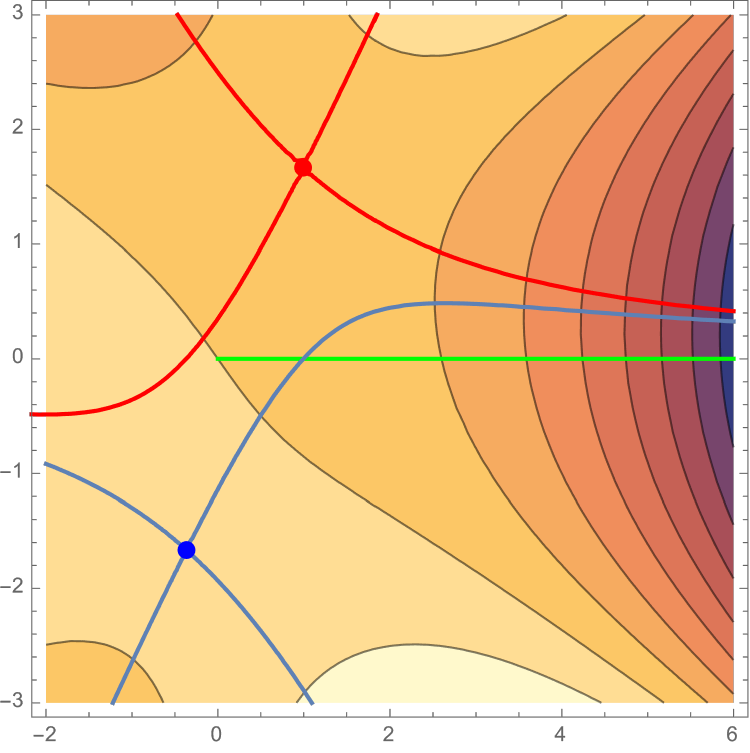}
        \caption{$\beta L = 40$}
    \end{subfigure}
    \caption{The panels show the real part of $u$ for $\mu_0 = 0.9$ and $\frac{n}{qL}=\frac{2}{3}$ using the same color coding as in Fig. \ref{fig:mu2}. The red steepest ascent line crosses the green contour in the second and third panels, but not in the first or last.  This illustrates that, for $qL = \frac{3}{2}n< \sqrt{3}n$ and $\mu_0<1$, this (${\mathcal R}_+$) saddle fails to contribute at large and small temperatures (respectively) for $\mu_0<1$, though it does contribute at intermediate temperatures.  The blue ($\mathcal{R}_-$) saddle never contributes since only its steepest {\it descent} line ever crosses the green contour.}
    \label{fig:mu09_n4}
\end{figure}

\begin{figure}[h!]
    \centering
    \includegraphics[width=0.35\textwidth]{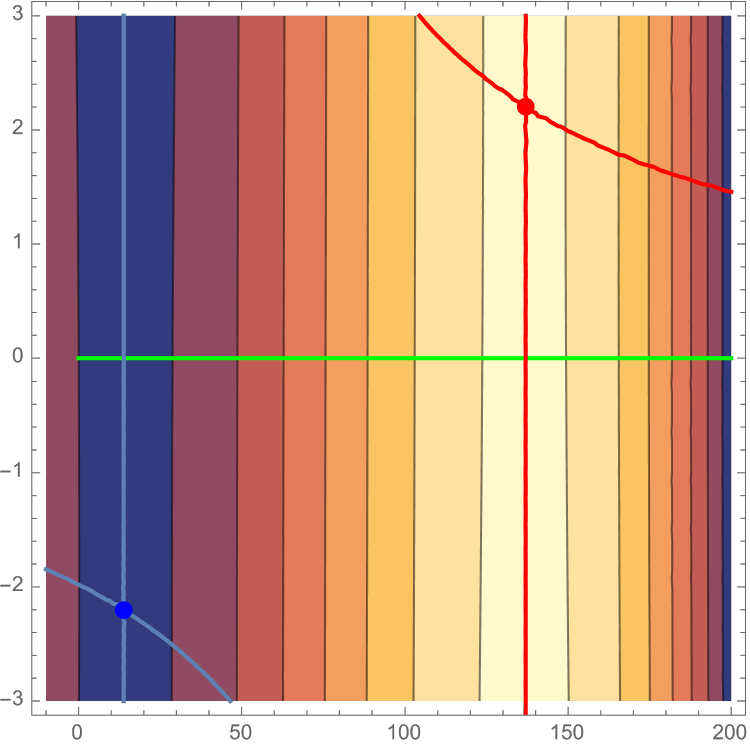}
    \includegraphics[width=0.35\textwidth]{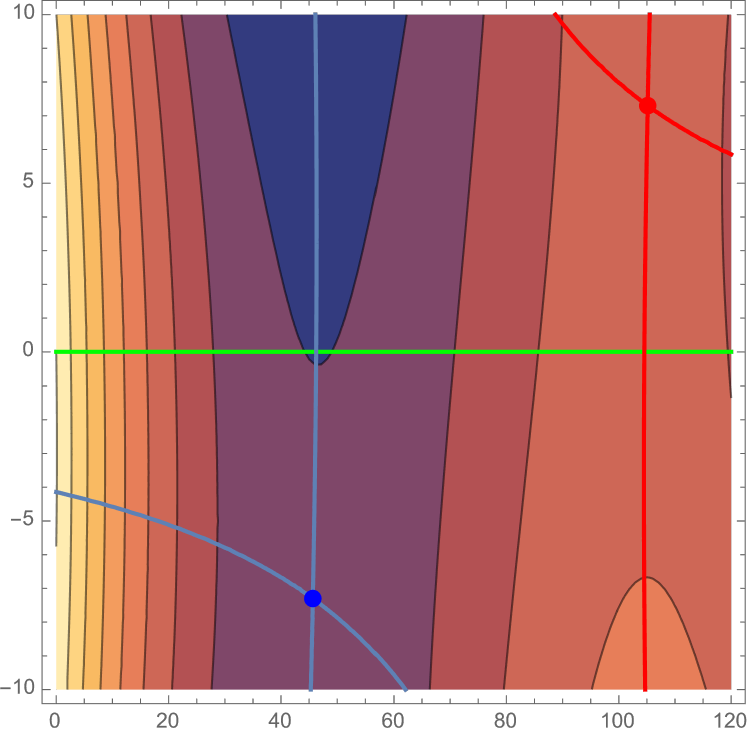}
    \caption{The figures show the real part of $u$ for $\mu_0=0.9$ at $\frac{\beta}{L}=1$ and $\frac{n}{qL}=\frac{1}{3}<\frac{1}{2}$ (left) and $\frac{n}{qL} = \frac{8}{15}>\frac{1}{2}$ (right),  using the same color coding as in Fig \ref{fig:mu2}.
    In both panels, the red steepest ascent line crosses the green contour.  Thus the red (${\mathcal R}_+$) saddle contributes in both cases. However, on the left it clearly gives a much larger contribution than the endpoint, while the endpoint contribution is much larger on the right. A blue line crosses the integration contour, but the shading indicates that it is a steepest descent contour, so the blue saddle fails to contribute in either case.}
    \label{fig:plot_Z_saddle_beta_1_mu09_n_different}
\end{figure}

\noindent
Figures \ref{fig:mu09_n3},
\ref{fig:mu09_n4} and \ref{fig:plot_Z_saddle_beta_1_mu09_n_different}
present examples of these behaviors.

\section{Pure Einstein-Hilbert gravity in AdS$_3$}
\label{sec:BTZ}
We can perform a similar analysis for uncharged (BTZ) black holes in $\textrm{AdS}_3$. The relevant action function is
\begin{equation}
S=     A/4 - \beta(E - \Omega J) = \frac{\pi \mathcal{R}}{2} - \beta \left(\frac{2J^2}{\mathcal{R}^2} + \frac{\mathcal{R}^2}{8} - \Omega J \right),
\end{equation}
where ${\mathcal R}= A/2\pi$.
The integral over $J$ is Gaussian and can be performed exactly to yield
\begin{equation}
Z_\Omega(\beta) :=     \sqrt{\frac{2\pi^3}{\beta}} \int_0^\infty \textrm{d}\mathcal{R} \mathcal{R} e^{\frac{1}{8} \mathcal{R} \left(\beta  \mathcal{R} \left(\Omega ^2-1\right)+4 \pi \right)}.
\label{eq:Zsubomega}
\end{equation}
Here, for convenience, we have defined $Z_\Omega$ so that it does {\it not} include the sum over shifted potentials and we will later compute
\begin{equation}
Z(\beta,\Omega) = \sum_{m\in {\mathbb Z}} Z_{\Omega+ \frac{2\pi m}{\beta s}i(\beta) }.
\end{equation}

Let us first note that \eqref{eq:Zsubomega} converges only for $\Re \Omega^2 < 1$.  This reflects the a standard condition in $\textrm{AdS}$ to avoid superradiant instability. When this condition holds, the result of the integral can be written exactly in terms of the error function. However, for studying the semiclassical contributions, it is more instructive to perform a saddle point and endpoint analysis in parallel with that in section \ref{sec:PL}.   

Since the exponent is quadratic in ${\mathcal R}$, there is a single saddle given by
\begin{equation}
    \mathcal{R}_\star = \frac{2\pi}{\beta (1-\Omega^2)}.
\end{equation}
At this value we find
\begin{equation}
    u(\mathcal{R}_\star) = \frac{\pi^2}{2\beta (1-\Omega^2)} =\frac{\pi}{4}\mathcal{R}_\star
\end{equation}
One may wonder why we did not find another saddle that (for real $\Omega$) would correspond to the inner horizon. The reason is just that, for BTZ black holes,  inner and outer horizons have non-overlapping ranges of $\Omega$.  In particular, while non-extreme outer horizons always have $\Omega<1,$ non-extreme inner horizons exist only with $\Omega >1$.

For real $\Omega< 1$, our saddle is clearly the maximum of the exponent and so  will always contribute. Writing $\Omega = \Omega_0 + \frac{2\pi i m}{\beta}$, we find
\begin{equation}
    u(\mathcal{R}_\star) = \frac{\pi^2}{2\beta \left(
    1 - \Omega_0^2 + \frac{4\pi^2 m^2}{\beta^2} \right) - 8 \Omega_0 \pi i m
    } = \frac{\pi^2 \left(2\beta \left(
    1 - \Omega_0^2 + \frac{4\pi^2 m^2}{\beta^2} \right) + 8 \Omega_0 \pi i m
    \right)}{4\beta^2 \left(
    1 - \Omega_0^2 + \frac{4\pi^2 m^2}{\beta^2} \right)^2 + 64 \Omega_0^2 \pi^2 m^2 }.
\end{equation}
On the other hand, writing $\mathcal{R}=a+b i$, we obtain
\begin{equation}
    \Im u(a+bi) = \frac{1}{4} \left(2 \pi  a^2 m \Omega_0+b \left(a \beta  \left(\Omega_0^2-1\right)-\frac{4 \pi ^2 a m^2}{\beta }+2 \pi \right)-2 \pi  b^2 m \Omega_0\right).
\end{equation}
We thus see that the curves of constant phase are generally hyperbolas. However, as is always the case for quadratic exponent functions, the constant phase curve that run through the saddle degenerate to straight lines. The steepest ascent and descent contours thus satisfy
\begin{equation}
    a = b \frac{4 \pi ^2 m^2-\beta ^2 \left(\Omega_0^2-1\right)}{4 \pi  \beta  m \Omega_0} \pm \frac{|b \left(\beta ^4 \left(\Omega_0^2-1\right)^2+16 \pi ^4 m^4+8 \pi ^2 \beta ^2 m^2 \left(\Omega_0^2+1\right)\right)-8 \pi ^2 \beta ^2 m \Omega_0|}{4 \pi  \beta  |m \Omega_0| \sqrt{\left(\beta ^2 (\Omega_0-1)^2+4 \pi ^2 m^2\right) \left(\beta ^2 (\Omega_0+1)^2+4 \pi ^2 m^2\right)}}.
\end{equation}

We immediately see that only the $+$ line crosses our integration contour (which has $a>0, b=0$). It remains only to determine whether the $+$ line is  ascending or descending as one moves away from the saddle.   It is easiest to check this by simply computing the function $u$ at the point where this contour crosses the positive real axis.  Doing so yields a value larger than at the saddle, so this curve is ascending. We thus conclude that our saddle always contributes.

Let us finish by checking what happens at large shifts.  We find
\begin{equation}
    u(\mathcal{R}_\star) = \frac{\beta }{8 m^2}+\frac{i \beta ^2 \Omega_0}{8 \pi  m^3} + O(m^{-4}),
\end{equation}
so that the action approach zero for either sign of $n$. Whether the sum over $n$ converges thus depends on the one-loop determinants. While we have certainly not included all one-loop effects in
\eqref{eq:Zsubomega}, the fact that $|\Omega^2-1|$ grows quadratically with $n$ at large $n$ means that, if we nevertheless treat \eqref{eq:Zsubomega} as exact, the corresponding sum of $Z_{\Omega + \frac{2\pi m}{\beta s}i }(\beta)$ over all $m$ does indeed converge.
We note that these saddles can be viewed as analytic continuations of the $M_{1,n}$ sub-family of $SL(2,\mathbb{Z})$ saddles of \cite{Maloney:2007ud}.

\section{Discussion}
\label{sec:disc}

The above work  used the Lorentzian contour prescriptions proposed in \cite{Marolf:2022ybi, Chen:2025leq} to compute the partition function of the spherically-symmetric sector of Einstein-Maxwell theory. We argued that it can be written as a sum over sectors associated with different values of the potential difference $\Delta \Phi$ between infinity and a codimenion-2 Lorentzian conical singularity, and where the relevant values of $\Delta \Phi$ are related by quantized imaginary shifts labeled by integers $n\in {\mathbb Z}$.   In general, each integral in this sum has many saddles in the complex plane and thus there are infinitely many saddles that could contribute to our partition function. Summing over all such saddles would render the answer divergent \cite{Harlow2022}. However, after using a constrained stationary-point approximation to reduce the path integrals to a finite-dimensional integral over a physically-motivated set of constrained saddles,  
a careful Picard-Lefschetz analysis shows that only a finite number of the above saddle contribute at any finite inverse temperature $\beta$.  This then resolves the issue raised in \cite{Harlow2022}. 
We also performed a corresponding analysis for pure Einstein-Hilbert gravity in AdS$_3$.  In that case we found that all saddles do in fact contribute, but that the sum nevertheless converges at the level  studied here.

Returning to the spherically-symmetric Einstein-Maxwell case, for all values of $\mu_0$ we find that only a finite number of saddles contribute at large temperature.  However, the behavior as we lower the temperature depends qualitatively on the chemical potential. The number of contributing saddles grows (roughly linearly) with $\beta$ for chemical potentials $|\mu_0| \ge 1$. In contrast, for $|\mu_0|<1$ we find that no saddle contributes in the limit $\beta \to \infty$.  Instead, in that limit each $Z_n$ receives only  endpoint contributions in the semiclassical limit.   Interestingly, the number of contributing saddles is not a monotonic function of the temperature. In particular, we found a regime where certain saddles contribute only in a finite interval of $\beta$ that is bounded away from zero. Importantly, in neither case is the correct result obtained by restricting to saddles in which our potential difference $\Delta \Phi$ between the horizon and infinity is purely real.

For completeness, appendix \ref{subsec:Euclidean} also studies partition functions of the form 
\begin{equation}
    \textrm{Tr} e^{-\beta (H - i \mu' Q)}
\end{equation}
for real $\mu' = -i\mu$.  For such cases, any configuration (with any value of $n$) will have purely imaginary $\Delta \Phi.$  Our results for which saddles contribute are in many ways qualitatively similar to those found above for real values of $\mu.$  However, in this case, and within the space of constrained saddles defined in section \ref{sec:PL}, all of the saddles that contribute to our path integral have real Euclidean geometries.  While complex saddles do arise for some values of $\mu',n$, the fact that the integrand is real in the relevant analogue of \eqref{eq:Zapprox2} means that such saddles can never contribute.  We emphasize that such a conclusion is far from clear if one attempts to take the Euclidean path integral as fundamental since, in that context, some non-real contour of integration must be chosen to resolve the conformal factor problem. 

Although we include only a finite number of spherically-symmetric Einstein-Maxwell saddles at each finite $\beta$, we emphasize that our partition functions are periodic under imaginary shifts in $\mu$ as desired.  The point here is that we still sum over sectors labeled by arbitrary values of $n \in {\mathbb Z}$.  It is only that we find that the contribution $Z_n$ from certain sectors is governed by an $O(1)$ endpoint contribution instead of by the expected saddle-point.  Furthermore, the set of saddles that do contribute depends on $\mu_0$.  Shifting $\mu_0$ by one (imaginary) period then also shifts the $n$-labels for the saddles that contribute, though the set of such saddle-contributions remains unchanged.

One-loop corrections are required for the sum over our endpoint  contributions to converge.  Here we have included one-loop contributions associated with our finite-dimensional integral over constrained saddles, though we have ignored other one-loop contributions.  Exploring the remaining one-loop effects, and perhaps justifying the above emphasis on those associated with our finite-dimensional integral over constrained saddles, remains a topic for future investigation.


Some readers may ask whether one can also use the Kontsevich-Segal-Witten (KSW) criterion \cite{Kontsevich:2021dmb, Witten:2021nzp} to address the validity of our complex saddles.  A complication, however, is that the application of KSW in such cases is not obviously well-defined.  In particular, the KSW criterion would allow a given complex metric to contribute  to the gravitational path integral (in the saddle point approximation) if it defines a convergent Euclidean path integral for matter fields\footnote{The use of this idea to pick out allowed complex metrics dates back at least to \cite{Halliwell:1988ik}, and was in particular used in \cite{Louko:1995jw}, which we have used heavily.  However, in \cite{Louko:1995jw} it suffices to use this condition in a context where all coordinates and matter fields are manifestly real.}.  In particular, it requires convergence of the matter path integral when integrated over {\it real} matter fields.  Since we may consider fluctuations of fields with compact support, one finds that  the matter action evaluated in an arbitrarily small region must have a positive real part, and -- assuming that one integrates over a set of real coordinates -- this analysis then restricts certain phases associated with the complex metric being considered.

However, in general it is far from clear that matter fluctuations about a given complex saddle should indeed be integrated over the real contour; see e.g. \cite{Marolf:2022ntb,Kolanowski:2024zrq} for recent examples.  Furthermore, such a  condition cannot be invariant under general complex coordinate transformations.  For example, multiplying each coordinate $x^a$ by $e^{i\phi}$ clearly multiplies the metric $g_{ab}$ by $e^{-2i\phi}$.  As a result, if a given metric satisfies KSW at a given real point $p$ (i.e., with real $x_p^a:= x^a(p)$, on the analytically extended geometry we can find other coordinates $y^a$ which happen to also be real at $p$, but which are generally related to $x^a$ by a complex coordinate transformation and for which the standard KSW condition is violated (and vice versa).  To do so, we need only take $y^a = e^{i\phi} (x^a-x^a_p) + O(x^a -x_p)^2$ for appropriate values of $\phi.$  Applying the KSW condition to the real section defined by taking real values of $x^a$ will thus generally give a different answer than that defined by taking real values of $y^a.$  This is clearly associated with the fact that, if a saddle has non-trivial matter fields, the matter fields (and any associated sense in which one would discuss fluctuations) will generally not be real valued on both of the above real sections; see e.g. \cite{Kolanowski:2024zrq} for a recent discussion of a simple example. Similar issues also appear to be related to violation of the KSW condition reported in  \cite{Chakravarty:2024bna} for the higher-dimensional double-cone (a saddle that is typically thought to be physically relevant).

Yet so long as we can deform the surface of real $x^a$ to the surface of real $y^a$ without passing through singularities of the Lagrangian, Cauchy's theorem guarantees that we find the same action by integrating the Lagrangian along either surface.  For this reason it is conventional to think of both choices as corresponding to the same complex saddle (as has been done implicitly in our work above).  Doing otherwise would lead to unphysical divergences from zero-modes associated with complex coordinate transformations. An explicit example in which it was shown that the one-loop determinant depends only on the homotopy class of the complex contour can be found in \cite{Turiaci:2025xwi}, see also earlier discussions in \cite{Witten:2021nzp}.  

While some complex saddles come with a natural set of coordinates that one can then restrict to real values\footnote{Examples include
Euclidean Kerr with imaginary angular momentum \cite{Witten:2021nzp}  or supersymmetric black holes in $\textrm{AdS}_5$ with the particular finely-tuned choices of parameters \cite{BenettiGenolini:2025jwe, BenettiGenolini:2026raa,Krishna:2026rma}. }, this is not the case for saddles at complex ${\mathcal R}_\pm$.  Instead, in order to apply KSW at all (at least while preserving spherical symmetry) we must first choose some contour that runs from ${\mathcal R}_\pm$ to $+\infty$ through the plane of complex ${\mathcal R}$, and on which we choose to use a real coordinate in terms of which we would then apply KSW.  

 Let us now work out an explicit example. We again consider  a complex AdS$_4$ Reissner-Nordstr\"om metric
\begin{subequations}
    \begin{equation}
    {\rm d}s^2 = f {\dd}t^2 + \frac{{\dd}r^2}{f} + r^2 \dd \Omega^2,
\end{equation}
where
\begin{equation}
    f(r) = \frac{r^2}{L^2}+1 -\frac{2m}{r} + \frac{Q^2}{r^2},
\end{equation}
and with the parameters $m$ and $Q$ fixed by choosing $\beta, \mu_n$.
\end{subequations}
For simplicity, we will restrict ourselves to piecewise straight contours in the complex $r$-plane. Let us write $r = x+iy$.  We first examine the lines of constant $x$ or constant $y$:
\begin{itemize}
    \item Horizontal line segment with $y$ constant: A short calculation shows the KSW criterion to be equivalent to requiring both $\ \Re\left(f(x+iy)\right) > 0$ and $x^2 > y^2$
    \item Vertical line segment with $x$ constant: We find that the KSW criterion is never satisfied.  
\end{itemize}
Given these results, we may expect more general straight lines to be allowed, provided that it is not too close to being vertical and that it lies at large enough values of $x$. Indeed, we have checked for various saddles that we expect to contribute that there exists an $\mathcal{R}_0 \in \mathbb{R}$ such that the KSW criterion is satisfied both on the interval connecting $\mathcal{R}_+$ with $\mathcal{R}_0$ as well on the half-line $[\mathcal{R}_0, \infty)$. Of course, the value of $\mathcal{R}_0$ is highly non-unique.   

However, we also found such $\mathcal{R}_0$ for complex saddles that do not contribute to our partition function.  A particularly significant example is Schwarzschild AdS$_4$ at $\beta \gtrsim \frac{2\sqrt{3}\pi}{3}L$ (the right-hand side being the critical $\beta$ above which there are no real Schwarzschild black holes). Even without our Lorentz-siganture motivations, one would be highly surprised to find that such black holes contribute to the desired partition function.  This example thus seems to imply that satisfying the KSW criterion along some contour cannot be a sufficient condition for a saddle to contribute. 

Interestingly, we found that for larger $\beta$ (i.e., for $\beta \gg \frac{2\sqrt{3}\pi}{3} L$) there is no such choice of $\mathcal{R}_0$ at all.  However, we have not investigated whether there might still be other more complicated contours on which the KSW criterion would be satisfied\footnote{It should be noted that these black holes have $\left(\Im \mathcal{R}_\pm \right)^2 > \left(\Re \mathcal{R}_\pm \right)^2$.}. Either way, these simple examples illustrate  the practical implementations of the KSW criterion in contexts with general complex saddles.

Returning to our Lorentzian approach, let us recall that we have explicitly included constrained saddles associated with inner-horizon values of $A,Q,J$ (see e.g. figure \ref{fig:inner} in section \ref{sec:PL}). It is an interesting question whether there might be some fundamental reason (or some loop effect associated with the closed or nearly-closed null geodesics mentioned below figure \ref{fig:inner}) why such constrained saddles should be excluded. If so, it would  decrease the size of our integration domain, potentially excluding additional saddles.  It would also change the endpoint contributions (or, more generally, boundary contributions). While we are not aware of any holographic computation that would be precise enough to be able to see corrections coming from such a change in the integration domain, it would nevertheless be useful to explore this further. 

Let us conclude by sketching possible extensions of the work presented above. So far, except for the BTZ case, we were happy to restrict ourselves to non-rotating solutions.  This was done largely because it simplified our analysis significantly, though also because the use of \eqref{eq:Zapprox2} has thus far been fully justified in the presence of rotation in 2+1 dimensions. However, we certainly expect that one may continue to use \eqref{eq:Zapprox2} in higher dimensions, and we look forward to investigations that better examine the desired generalization of \cite{Chen:2025leq}.  

It would also be interesting to investigate contexts where higher derivative terms play an important role.
While there is certainly more to understand regarding theories with such terms,  it was proposed in \cite{Colin-Ellerin:2020mva} to define the action using the Legendre transform of the action described in appendix B of \cite{Dong:2019piw}.

Now, as mentioned in section \ref{sec:PL}, it is tempting to argue solely on physical grounds that 
that \eqref{eq:Zapprox2} is a good approximation
even in contexts where  a derivation from the Lorentzian path integral is not yet in hand.  If one does so, it is clear that one will find qualitatively similar structure regarding the relevance of AdS Kerr-Newman saddles.  In particular, the inequality \eqref{eq:Z00} will again suffice to exclude a large class of rotating saddles, though it cannot tell us which saddles actually contribute.  Similar issues also  arise naturally when studying a supersymmetric index \cite{Chen:2023mbc}, a spin-refined partition function \cite{Grabovsky:2024vnb},  or  general partition functions \cite{Iliesiu:2026}.
An especially interesting case involves computing the $\frac{1}{16}$-BPS index for supergravity on $\textrm{AdS}_5\times S^5$ (and which is dual to $\mathcal{N}=4$ super Yang-Mills), which will be explored using our methods in forthcoming work \cite{Kolanowski:2026aaa}.

\section*{Acknowledgements}
We thank Joaquin Turiaci for conversations that motivated this work. We also thank Vincent Chen and Luca Iliesiu for interesting related discussions.  This work was supported by NSF grant PHY-2408110 and by funds from the University of California. This
work was performed in part at the Aspen Center for Physics, which is supported by a grant
from the Simons Foundation (1161654, Troyer)

\appendix
    
\section{Analytic form of the Lefschetz thimbles} \label{app:analytic}

We will now provide analytic expressions for the constant phase curves.

Let us start by observing that the function $u$ can be written as
\begin{subequations}
\begin{equation}
    u(\mathcal{R}) = ({\mathcal R}-{\mathcal R}_\pm)^2 (x (\mathcal{R}-\mathcal{R}_\pm) + y) + u(R_\pm),
\end{equation}
where 
\begin{equation}
    x =-\frac{\beta}{2L^2}=|x| e^{i \pi}
\end{equation}
and
\begin{equation}
    y = \mp\frac{\sqrt{3 \beta ^2 \left(\mu_n ^2-1\right)+4 \pi ^2 L^2}}{2 L}=|y| e^{i \alpha_2},
\end{equation}
\end{subequations}
The constant phase curves that go through the critical point $({\mathcal R}_\pm, u({\mathcal R}_\pm))$ are simply those for which $({\mathcal R}-{\mathcal R}_\pm)^2 (x (\mathcal{R}-\mathcal{R}_\pm) + y)$ is purely real. Writing
\begin{equation}
    \mathcal{R} = \mathcal{R}_\pm + r e^{i\phi},
\end{equation}
with $r \ge 0$ and $\phi \in [0,2\pi)$, we see that this condition is equivalent to
\begin{equation}
    0 = \Im \left(
({\mathcal R}-{\mathcal R}_\pm)^2 (x (\mathcal{R}-\mathcal{R}_\pm) + y)
    \right) = r^2 \left(-
|x| r \sin \left(3\phi\right) + |y| \sin \left(2\phi + \alpha_2 \right)
    \right).
\end{equation}
The curve is thus given by
\begin{equation}
    r = \frac{|y|}{|x|} \frac{\sin \left(2\phi + \alpha_2 \right)}{\sin \left(3\phi \right)},
\end{equation}
and, of course, $r=0$. This representation makes sense only for values of $\phi$ such that $\frac{\sin \left(2\phi + \alpha_2 \right)}{\sin \left(3\phi \right)} \ge 0$.  The question of such a curve crossing our contour (the positive real axis) reduces to the following equation
\begin{subequations}
    \begin{equation}
        \Im \mathcal{R}_\pm  + \frac{|y|}{|x|} \frac{\sin \left(2\phi + \alpha_2 \right)}{\sin \left(3\phi\right)} \sin(\phi) =0
    \end{equation}
   together with the constraint
    \begin{equation}
        \Re \mathcal{R}_\pm  + \frac{|y|}{|x|} \frac{\sin \left(2\phi + \alpha_2 \right)}{\sin \left(3\phi \right)} \cos(\phi) >0.
    \end{equation}
\end{subequations}
One can show that the latter is equivalent to the imaginary part of the action at the critical point having the same sign as the imaginary part of the action along our contour (which, of course, is a necessary condition for the constant phase line to cross the contour).

\begin{figure}[t]
    \centering
    \includegraphics[width=\textwidth]{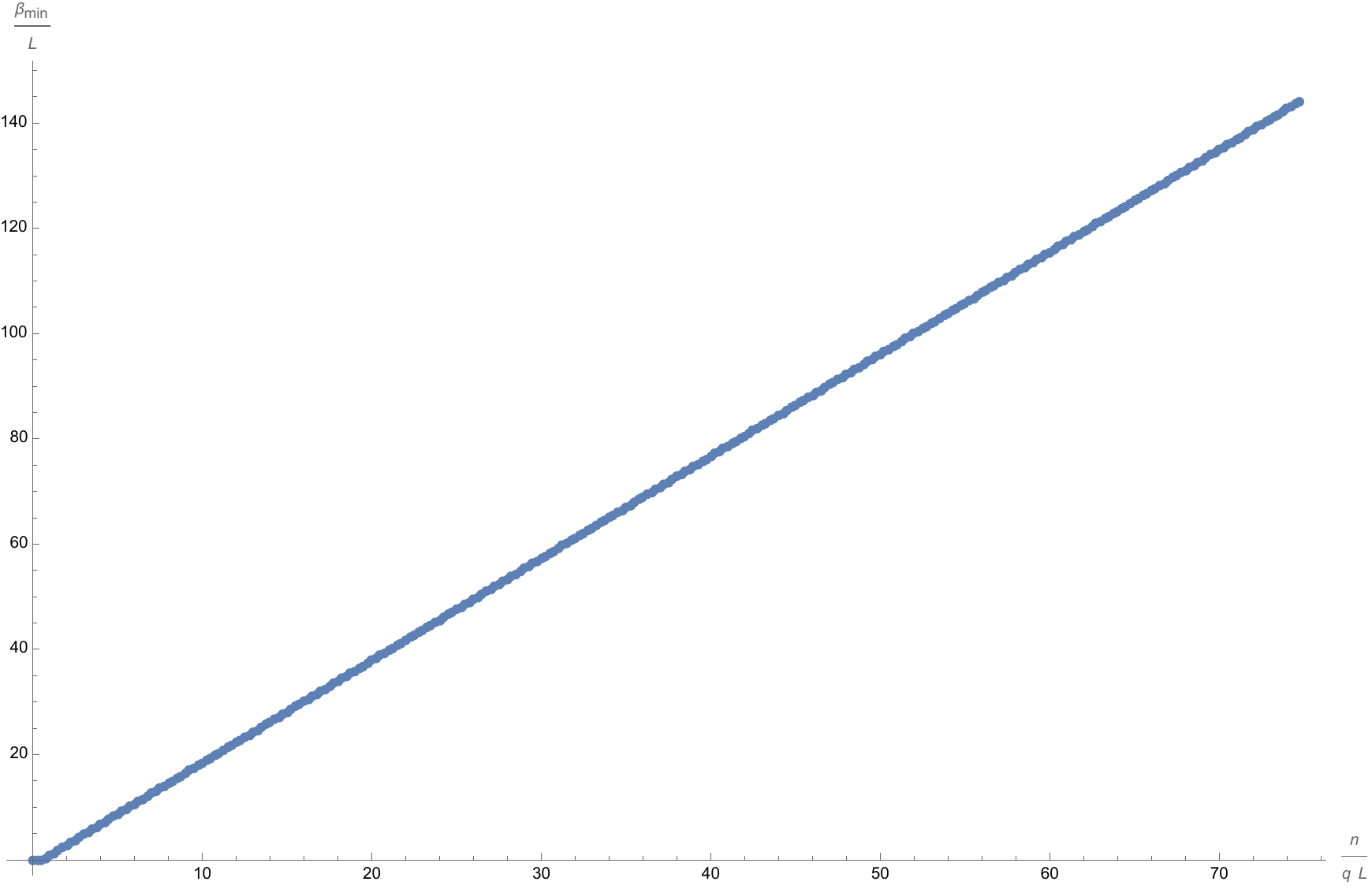}
\caption{The  dependence of the minimal inverse temperature $\beta$ at which a saddle contributes as a function of the number of shifts $n$ made in the imaginary direction for $\mu_0 = 2$. Notice that for $\frac{n}{qL}< \frac{1}{\sqrt{3}}$, saddles contribute at all temperatures.}
    \label{fig:beta_min}
\end{figure}

The first equation can be solved in a compact way if we notice that
\begin{equation}
    \mathcal{R}_\pm = \frac{\pi}{3|x|} - \frac{|y|}{3|x|} e^{i\alpha_2}.
\end{equation}
This immediately gives
\begin{equation}
    \tan \phi = \frac{3}{\tan \alpha_2}.
\end{equation}
Solutions to this equation are defined up to integer shifts by $\pi$. Since $\phi \in [0,2\pi)$, we only need to consider two solutions that differ by $\pi$.  We must always choose the one that gives $r>0$. It is easy to see that it defines $\phi$ uniquely. However, it does not appear possible to analytically to determine if this value satisfies the constraint since the condition reduces to a transcendent inequality. Nevertheless, we may solve numerically for the minimal $\beta$ for which the constraint is satisfied. Quite surprisingly, we found a very linear trend (see Fig. \ref{fig:beta_min}). However, one may also show that the apparently-linear dependence of $\beta_{\textrm{min}}$ on $n$ is exact. In particular, at large $n$ we find:
\begin{equation}
     \beta_{\textrm{min}} = \frac{2 \pi  \left(\sqrt{4 \mu_0^2-3}-\mu_0\right)}{\sqrt{3} \left(\mu_0^2-1\right)} \frac{|n|}{q} -\frac{2 \pi  L \mu_0}{\sqrt{\mu_0 \left(4 \mu_0^2-3\right) \left(\sqrt{4 \mu_0^2-3}+\mu_0\right)}} + O \left(
     \frac{q}{|n|}
     \right).
\end{equation}
This result confirms that, at large $n$, the saddle starts contributing when $q\beta$ is of order $n$.

However, let us now consider an opposite regime  in which $|n|$ is (relatively) small, and in particular with $Lq > \sqrt{3}|n|$. For brevity, let us write $x = \frac{n}{q}$ so that the condition reads $L> \sqrt{3}|x|$. Evaluating $u(\mathcal{R}_+)$ at large temperatures (small $\beta$) shows that
\begin{equation}
    u\left(\mathcal{R}_+\right) = \frac{4 \pi ^3 L}{27 \beta ^2} \left(2 L^3+2 \left(L^2 - 3x^2\right)^{3/2}-9 L x^2\right) + \frac{4 i \pi ^2 L \mu_0 x \left(\sqrt{L^2-3x^2}+L\right)}{3 \beta} +O\left(\beta^0\right).
    \label{eq:smallbeta}
\end{equation}
Importantly, as long as $L^2 > 3x^2$, the first term is real and the second term is purely imaginary. The latter has the right sign for the saddle to potentially contribute. A direct numerical analysis of the contours  then shows that it indeed contributes for all $\beta$ (there is no Stokes phenomenon). Furthermore, if $L^2 > 4x^2$ the first term in \eqref{eq:smallbeta} is positive. As a result, at large temperatures the saddle contribution is very large and dominates over the order-one endpoint contribution. On the other hand, for $4x^2>L^2 > 3x^2$, the saddle contributes but, at small $\beta$, it is subdominant with respect to the boundary contribution.

\section{Real Euclidean boundary conditions}
\label{subsec:Euclidean}

For completeness, we will now consider a path integral computation of
\begin{equation}
    \textrm{Tr} e^{-\beta (H - i \mu' Q)}
\end{equation}
for real $\mu'$; i.e., we now take $\mu=i\mu'$ to be imaginary.  As remarked in the introduction, in this case we can take the boundary conditions of the Euclidean path integral to be real for all fields.  Of course, due to the conformal factor problem one still cannot take the full Euclidean path integral to be defined in terms of the real contour.  But one might expect at least certain simplification in the analysis and, indeed, this is the case at the level of applying the general result \eqref{eq:Zapprox2} (which, as noted in section \ref{sec:chsing}, follows equally well from our Lorentian path integral prescription for any complex $\mu$).

In the present case, \eqref{eq:Zapprox2} again becomes \eqref{eq:Zsum}, which we may write in the form
\begin{equation}
Z(\beta, i\mu'):=    \sum_{n \in \mathbb{Z}} \int dA dQ e^{A/4}e^{-\beta  \left(E - i\left(\mu' + \frac{2\pi n}{\beta} \right)Q \right)} =: \sum_{n \in \mathbb{Z}} Z_n.
\end{equation}
Let us now denote $\mu_n{}' = \mu' + \frac{2\pi n}{\beta}$. The integral over $Q$ remains gaussian and gives
    \begin{align}
\begin{split}
    Z_{n} &= 8 \pi \int d\mathcal{R} \ \mathcal{R} e^{\pi \mathcal{R}^2 - \beta \left(
    \frac{\mathcal{R}^3}{2L^2} + \frac{\mathcal{R}}{2}
    \right)} \int dQ e^{- \beta\left(
\frac{Q^2}{2\mathcal{R}} - i\mu_n{}' Q
    \right)}\\ &= 8 \sqrt{2} \pi^{3/2} \beta^{-1/2} \int d\mathcal{R} \mathcal{R}^{3/2} \exp \left(
- \beta \left(
\frac{\mathcal{R}^3}{2L^2} + \frac{\mathcal{R}}{2} \left(
1+(\mu_n{}')^2
\right)
\right) + \pi \mathcal{R}^2
    \right).
\end{split}
\end{align}
In particular, the exponent is 
\begin{equation}
    u(\mathcal{R}) = - \beta \left(
\frac{\mathcal{R}^3}{2L^2} + \frac{\mathcal{R}}{2} \left(
1+(\mu_n{}')^2
\right)
\right) + \pi \mathcal{R}^2.
\end{equation}

In this case,  since $\mu_n \in \mathbb{R}$, we are performing a purely real integral along the real half-line.  As such, it can only receive contributions from critical points lying on the original integration contour (i.e., from real Euclidean black holes), and in particular from those that are local maxima of the integrand along this contour; see again footnote \ref{foot:realint} from section \ref{sec:PL}.   The possible saddles are just  analytic continuations of \eqref{ref:saddles} and so yield
\begin{subequations}
\begin{equation}
    \mathcal{R}_\pm = \frac{L \left(2\pi L \pm \sqrt{4 \pi ^2 L^2-3 \beta ^2 \left((\mu_n{}') ^2+1\right)}\right)}{3 \beta } 
    \end{equation}
    \begin{eqnarray}
    u(\mathcal{R}_\pm) &=& \frac{L}{27 \beta ^2} \left(\sqrt{4 \pi ^2 L^2-3 \beta ^2 \left((\mu_n{}') ^2+1\right)} \pm 2 \pi  L\right)  \\ &\times& \left(\pi  L \sqrt{4 \pi ^2 L^2-3 \beta ^2 \left((\mu_n{}') ^2+1\right)}\pm2 \pi ^2 L^2 \mp 3 \beta ^2 \left((\mu_n{}') ^2+1\right)\right).
    \end{eqnarray}
\end{subequations}
Since $u\rightarrow -\infty$ as ${\mathcal R}\rightarrow +\infty$, when ${\mathcal R}_\pm$ are real we see that ${\mathcal R}_+$ is a local maximum while ${\mathcal R}_-$ is a local minimum. Thus only the former can contribute. 
 This sort of phenomenon may in part explain why perturbative approaches have tended to yield physically-reasonable results in past studies of real Euclidean saddles; see e.g.
\cite{Gibbons:1978ac,Garriga:1997wz,Prestidge:1999uq,Gratton:1999ya,Kol:2006ga,Marolf:2022ntb, Liu:2023jvm}.

We also immediately see that these critical points are real only if 
\begin{equation}
    4\pi^2 L^2 > 3 \beta^2 \left((\mu_n{}')^2 +1 \right).
\end{equation}
In particular, at low temperature (large $\beta$), the saddles are complex for any $\mu$.  This should not come as a surprise since we know that real Euclidean Schwarzschild AdS black holes fail to exist at low temperatures.  Furthermore, since $\mu_n{}' = \mu_0{}' + \frac{2\pi n}{q \beta}$, at fixed real $\mu_0{}'$ and for sufficiently large $|n|$ there is no value of $\beta$ for which the saddle is real. For example, for $\mu_0=0$, only saddles with $Lq \ge \sqrt{3}|n|$ are real. All other $Z_{n \neq 0}$ are then dominated by the endpoint contribution and they are of order $1$.

The results for real $\mu_0{}'$ (imaginary $\mu_0$) are thus qualitatively similar to those for real $\mu_0$, at least at large $n$.  This should not be a surprise since, in both cases, moving to large $n$ corresponds to taking the imaginary part of $\mu_n$ large.  In particular, while taking $\mu_0$ imaginary does not automatically make saddles of $u$ real, when combined with our Lorentzian prescription, at least at the current level of approximation it {\it does} guarantee that only saddles with real ${\mathcal R}$ can contribute to the partition function.

\bibliographystyle{JHEP}
\bibliography{bibl.bib}

@article{Marolf:2022ybi,
    author = "Marolf, Donald",
    title = "{Gravitational thermodynamics without the conformal factor problem: partition functions and Euclidean saddles from Lorentzian path integrals}",
    eprint = "2203.07421",
    archivePrefix = "arXiv",
    primaryClass = "hep-th",
    doi = "10.1007/JHEP07(2022)108",
    journal = "JHEP",
    volume = "07",
    pages = "108",
    year = "2022"
}

@article{Saad:2019lba,
    author = "Saad, Phil and Shenker, Stephen H. and Stanford, Douglas",
    title = "{JT gravity as a matrix integral}",
    eprint = "1903.11115",
    archivePrefix = "arXiv",
    primaryClass = "hep-th",
    month = "3",
    year = "2019"
}

@article{Stokes:1847,
    author = "Stokes, G. G.",
    title = "{On the numerical calculation of a class of definite integrals and infinite series}",
    journal = "Transactions of the Cambridge Philosophical Society",
    volume = "IX (I)",
    pages = "166-189",
    year = "1847"
}

@article{Halliwell:1989dy,
    author = "Halliwell, Jonathan J. and Hartle, James B.",
    title = "{Integration Contours for the No Boundary Wave Function of the Universe}",
    reportNumber = "NSF-ITP-89-147",
    doi = "10.1103/PhysRevD.41.1815",
    journal = "Phys. Rev. D",
    volume = "41",
    pages = "1815",
    year = "1990"
}

@article{Halliwell:1988ik,
    author = "Halliwell, Jonathan J. and Louko, Jorma",
    title = "{Steepest Descent Contours in the Path Integral Approach to Quantum Cosmology. 1. The De Sitter Minisuperspace Model}",
    reportNumber = "NSF-ITP-88-173",
    doi = "10.1103/PhysRevD.39.2206",
    journal = "Phys. Rev. D",
    volume = "39",
    pages = "2206",
    year = "1989"
}

@article{Held:2024qcl,
    author = "Held, Jesse and Liu, Xiaoyi and Marolf, Donald and Wang, Zhencheng",
    title = "{Euclidean and complex geometries from real-time computations of gravitational R{\'e}nyi entropies}",
    eprint = "2409.17428",
    archivePrefix = "arXiv",
    primaryClass = "hep-th",
    doi = "10.1007/JHEP02(2025)136",
    journal = "JHEP",
    volume = "02",
    pages = "136",
    year = "2025"
}

@article{FAs,
author = "M. V. Fedorjuk",
title = "The Asymptotics Of The Fourier Transform Of The Exponential Function Of A Polynomial",
journal = "Soviet Math. Doklady",
volume = "17",
year = "1976",
pages = "486-490"
}

@article{FP,
author = "Fr\'ed\'eric Pham",
title ="Vanishing Homologies And The $n$ Variable Saddlepoint Method",
journal = "Proc. Symp. Pure Math.",
issue = "40",
volume = "2",
year = "1983",
page =  "319-333"
}

@book{AGV,
author = "V. I. Arnold and S. M. Gusein-Zade and A. N. Varchenko",
title = "Singularities Of Differentiable Maps, Vol. 2",
publisher = "Birkhauser",
year =  "1988"
}

@article{BH,
author = "M. V. Berry and C. J. Howls",
title = "Hyperasymptotics",
journal = "Proc. R. Soc.",
volume =  "A430",
pages =  "653-668",
year = "1990"
}

@article{BH2,
author =  "M. V. Berry and C. J. Howls",
title = "Hyperasymptotics For Integrals With Saddles",
journal = "Proc. R. Soc.",
volume = "A434",
pages = "657-675",
year = "1991"
}

@article{H,
author = "C. J. Howls",
title = "Hyperasymptotics For Multidimensional Integrals, Exact Remainder Terms And The Global Connection Problem",
journal = "Proc. R. Soc. London",
volume = "453",
year = "1997",
pages =  "2271"
}

@article{Marolf:2020rpm,
    author = "Marolf, Donald and Maxfield, Henry",
    title = "{Observations of Hawking radiation: the Page curve and baby universes}",
    eprint = "2010.06602",
    archivePrefix = "arXiv",
    primaryClass = "hep-th",
    doi = "10.1007/JHEP04(2021)272",
    journal = "JHEP",
    volume = "04",
    pages = "272",
    year = "2021"}

@article{Colin-Ellerin:2020mva,
    author = "Colin-Ellerin, Sean and Dong, Xi and Marolf, Donald and Rangamani, Mukund and Wang, Zhencheng",
    title = "{Real-time gravitational replicas: Formalism and a variational principle}",
    eprint = "2012.00828",
    archivePrefix = "arXiv",
    primaryClass = "hep-th",
    doi = "10.1007/JHEP05(2021)117",
    journal = "JHEP",
    volume = "05",
    pages = "117",
    year = "2021"
}

@article{Colin-Ellerin:2021jev,
    author = "Colin-Ellerin, Sean and Dong, Xi and Marolf, Donald and Rangamani, Mukund and Wang, Zhencheng",
    title = "{Real-time gravitational replicas: low dimensional examples}",
    eprint = "2105.07002",
    archivePrefix = "arXiv",
    primaryClass = "hep-th",
    doi = "10.1007/JHEP08(2021)171",
    journal = "JHEP",
    volume = "08",
    pages = "171",
    year = "2021"
}

@inproceedings{Hartle:2020glw,
    title = "{The Conformal Rotation in Linearised Gravity}",
    author = "Hartle, James B. and Schleich, Kristen",
    booktitle="Quantum Field Theory and Quantum Statistics",
    editor="I. A. Batalin, C.
J. Isham and G. A. Vilkovisky",
published="Adam Hilger",
pages = "{67-87}",
    eprint = "2004.06635",
    archivePrefix = "arXiv",
    primaryClass = "gr-qc",
    month = "4",
    year = "1987"
}

@article{Marolf:1996gb,
    author = "Marolf, Donald",
    title = "{Path integrals and instantons in quantum gravity: Minisuperspace models}",
    eprint = "gr-qc/9602019",
    archivePrefix = "arXiv",
    reportNumber = "UCSBTH-96-01",
    doi = "10.1103/PhysRevD.53.6979",
    journal = "Phys. Rev. D",
    volume = "53",
    pages = "6979--6990",
    year = "1996"
}

@inproceedings{Giddings:1990yj,
    author = "Giddings, Steven B.",
    title = "{Wormholes, the conformal factor, and the cosmological constant}",
    booktitle = "{International Colloquium on Modern Quantum Field Theory}",
    reportNumber = "HUTP-90-A010",
    month = "5",
    year = "1990"}

@article{Giddings:1989ny,
    author = "Giddings, Steven B.",
    title = "{The Conformal Factor and the Cosmological Constant}",
    reportNumber = "HUTP-89/A056",
    doi = "10.1142/S0217751X9000163X",
    journal = "Int. J. Mod. Phys. A",
    volume = "5",
    pages = "3811--3830",
    year = "1990"
}

@article{Schleich:1987fm,
    author = "Schleich, K.",
    editor = "Gibbons, G. W. and Hawking, S. W.",
    title = "{Conformal Rotation in Perturbative Gravity}",
    doi = "10.1103/PhysRevD.36.2342",
    journal = "Phys. Rev. D",
    volume = "36",
    pages = "2342--2363",
    year = "1987"
}

@article{Mazur:1989by,
    author = "Mazur, Pawel O. and Mottola, Emil",
    title = "{The Gravitational Measure, Solution of the Conformal Factor Problem and Stability of the Ground State of Quantum Gravity}",
    reportNumber = "UFIFT-AST-89-3, LA-UR-89-340",
    doi = "10.1016/0550-3213(90)90268-I",
    journal = "Nucl. Phys. B",
    volume = "341",
    pages = "187--212",
    year = "1990"
}

@article{Dasgupta:2001ue,
    author = "Dasgupta, A. and Loll, R.",
    title = "{A Proper time cure for the conformal sickness in quantum gravity}",
    eprint = "hep-th/0103186",
    archivePrefix = "arXiv",
    reportNumber = "AEI-2001-020",
    doi = "10.1016/S0550-3213(01)00227-9",
    journal = "Nucl. Phys. B",
    volume = "606",
    pages = "357--379",
    year = "2001"
}

@article{Ambjorn:2002gr,
    author = "Ambjorn, J. and Dasgupta, A. and Jurkiewicz, J. and Loll, R.",
    title = "{A Lorentzian cure for Euclidean troubles}",
    eprint = "hep-th/0201104",
    archivePrefix = "arXiv",
    reportNumber = "SPIN-2001-33",
    doi = "10.1016/S0920-5632(01)01903-X",
    journal = "Nucl. Phys. B Proc. Suppl.",
    volume = "106",
    pages = "977--979",
    year = "2002"
}

@article{Dong:2016hjy,
    author = "Dong, Xi and Lewkowycz, Aitor and Rangamani, Mukund",
    title = "{Deriving covariant holographic entanglement}",
    eprint = "1607.07506",
    archivePrefix = "arXiv",
    primaryClass = "hep-th",
    doi = "10.1007/JHEP11(2016)028",
    journal = "JHEP",
    volume = "11",
    pages = "028",
    year = "2016"
}

@article{Neiman:2013ap,
    author = "Neiman, Yasha",
    title = "{The imaginary part of the gravity action and black hole entropy}",
    eprint = "1301.7041",
    archivePrefix = "arXiv",
    primaryClass = "gr-qc",
    reportNumber = "IGC-13-1-1",
    doi = "10.1007/JHEP04(2013)071",
    journal = "JHEP",
    volume = "04",
    pages = "071",
    year = "2013"
}

@article{Halliwell:1989vu,
    author = "Halliwell, Jonathan J. and Louko, Jorma",
    title = "{Steepest Descent Contours in the Path Integral Approach to Quantum Cosmology. 2. Microsuperspace}",
    reportNumber = "NSF-ITP-89-21",
    doi = "10.1103/PhysRevD.40.1868",
    journal = "Phys. Rev. D",
    volume = "40",
    pages = "1868",
    year = "1989"
}

@article{Halliwell:1990tu,
    author = "Halliwell, Jonathan J. and Louko, Jorma",
    title = "{Steepest Descent Contours in the Path Integral Approach to Quantum Cosmology. 3. A General Method With Applications to Anisotropic Minisuperspace Models}",
    reportNumber = "MIT-CTP-1846",
    doi = "10.1103/PhysRevD.42.3997",
    journal = "Phys. Rev. D",
    volume = "42",
    pages = "3997--4031",
    year = "1990"
}

@article{Stokes:1858,
    author = "Stokes, G. G.",
    title = "{On the discontinuity of arbitrary constants which appear in divergent developments}",
    journal = "Transactions of the Cambridge Philosophical Society",
    volume = "X (I)",
    pages = "105-128",
    year = "1858"
}

@article{Witten:2010cx,
    author = "Witten, Edward",
    editor = "Andersen, Joergen E. and Boden, Hans U. and Hahn, Atle and Himpel, Benjamin",
    title = "{Analytic Continuation Of Chern-Simons Theory}",
    eprint = "1001.2933",
    archivePrefix = "arXiv",
    primaryClass = "hep-th",
    journal = "AMS/IP Stud. Adv. Math.",
    volume = "50",
    pages = "347--446",
    year = "2011"
}

@article{Chen:2025leq,
    author = "Chen, Hong Zhe",
    title = "{Thermodynamic stability from Lorentzian path integrals and codimension-two singularities}",
    eprint = "2501.08409",
    archivePrefix = "arXiv",
    primaryClass = "hep-th",
    doi = "10.1007/JHEP05(2025)180",
    journal = "JHEP",
    volume = "05",
    pages = "180",
    year = "2025"
}

@unpublished{Harlow2022,
  author       = {Harlow, D. and Iliesiu, L. and Ooguri, H. and Turiaci, J.},
  title        = {},
  note         = {Unpublished notes},
  year         = {2022}
}

@article{Heydeman:2020hhw,
    author = "Heydeman, Matthew and Iliesiu, Luca V. and Turiaci, Gustavo J. and Zhao, Wenli",
    title = "{The statistical mechanics of near-BPS black holes}",
    eprint = "2011.01953",
    archivePrefix = "arXiv",
    primaryClass = "hep-th",
    reportNumber = "PUPT-2621",
    doi = "10.1088/1751-8121/ac3be9",
    journal = "J. Phys. A",
    volume = "55",
    number = "1",
    pages = "014004",
    year = "2022"
}

@article{Kolanowski:2024zrq,
    author = "Kolanowski, Maciej and Marolf, Donald and Rakic, Ilija and Rangamani, Mukund and Turiaci, Gustavo J.",
    title = "{Looking at extremal black holes from very far away}",
    eprint = "2409.16248",
    archivePrefix = "arXiv",
    primaryClass = "hep-th",
    doi = "10.1007/JHEP04(2025)020",
    journal = "JHEP",
    volume = "04",
    pages = "020",
    year = "2025"
}

@article{Marolf:2022ntb,
    author = "Marolf, Donald and Santos, Jorge E.",
    title = "{The canonical ensemble reloaded: the complex-stability of Euclidean quantum gravity for black holes in a box}",
    eprint = "2202.11786",
    archivePrefix = "arXiv",
    primaryClass = "hep-th",
    doi = "10.1007/JHEP08(2022)215",
    journal = "JHEP",
    volume = "08",
    pages = "215",
    year = "2022"
}

@article{Liu:2023jvm,
    author = "Liu, Xiaoyi and Marolf, Donald and Santos, Jorge E.",
    title = "{Stability of saddles and choices of contour in the Euclidean path integral for linearized gravity: dependence on the DeWitt parameter}",
    eprint = "2310.08555",
    archivePrefix = "arXiv",
    primaryClass = "hep-th",
    doi = "10.1007/JHEP05(2024)087",
    journal = "JHEP",
    volume = "05",
    pages = "087",
    year = "2024"
}

@article{Gibbons:1978ac,
    author = "Gibbons, G. W. and Hawking, S. W. and Perry, M. J.",
    title = "{Path Integrals and the Indefiniteness of the Gravitational Action}",
    reportNumber = "PRINT-78-0375 (CAMBRIDGE)",
    doi = "10.1016/0550-3213(78)90161-X",
    journal = "Nucl. Phys. B",
    volume = "138",
    pages = "141--150",
    year = "1978"
}

@article{Hawking:1982dh,
    author = "Hawking, S. W. and Page, Don N.",
    title = "{Thermodynamics of Black Holes in anti-De Sitter Space}",
    reportNumber = "PRINT-83-0019 (CAMBRIDGE)",
    doi = "10.1007/BF01208266",
    journal = "Commun. Math. Phys.",
    volume = "87",
    pages = "577",
    year = "1983"
}

@article{Mahajan:2025bzo,
    author = "Mahajan, Raghu and Singhi, Kaustubh",
    title = "{A brief note on complex AdS-Schwarzschild black holes}",
    eprint = "2509.08883",
    archivePrefix = "arXiv",
    primaryClass = "hep-th",
    doi = "10.1007/JHEP11(2025)164",
    journal = "JHEP",
    volume = "11",
    pages = "164",
    year = "2025"
}

@article{Singhi:2025rfy,
    author = "Singhi, Kaustubh",
    title = "{Complex Kerr-AdS Black Holes}",
    eprint = "2510.01313",
    archivePrefix = "arXiv",
    primaryClass = "hep-th",
    month = "10",
    year = "2025"
}

@article{Ailiga:2025osa,
    author = "Ailiga, Manishankar and Mallik, Shubhashis and Narain, Gaurav",
    title = "{Complex saddles of charged-AdS gravitational partition function}",
    eprint = "2510.25396",
    archivePrefix = "arXiv",
    primaryClass = "hep-th",
    doi = "10.1007/JHEP02(2026)054",
    journal = "JHEP",
    volume = "02",
    pages = "054",
    year = "2026"
}

@article{Held:2026huj,
    author = "Held, Jesse and Kaplan, Molly and Marolf, Donald and Wang, Zhencheng",
    title = "{Axion Wormholes and the AdS/CFT Factorization Problem}",
    eprint = "2601.02507",
    archivePrefix = "arXiv",
    primaryClass = "hep-th",
    month = "1",
    year = "2026"
}

@article{Held:2026bbo,
    author = "Held, Jesse and Kaplan, Molly and Marolf, Donald and Wang, Zhencheng",
    title = "{Lorentzian Path Integrals and Jackiw-Teitelboim wormholes with imaginary scalars}",
    eprint = "2601.09932",
    archivePrefix = "arXiv",
    primaryClass = "hep-th",
    month = "1",
    year = "2026"
}

@article{Feldbrugge:2017kzv,
    author = "Feldbrugge, Job and Lehners, Jean-Luc and Turok, Neil",
    title = "{Lorentzian Quantum Cosmology}",
    eprint = "1703.02076",
    archivePrefix = "arXiv",
    primaryClass = "hep-th",
    doi = "10.1103/PhysRevD.95.103508",
    journal = "Phys. Rev. D",
    volume = "95",
    number = "10",
    pages = "103508",
    year = "2017"
}

@article{Gratton:1999ya,
    author = "Gratton, Steven and Turok, Neil",
    title = "{Cosmological perturbations from the no boundary Euclidean path integral}",
    eprint = "astro-ph/9902265",
    archivePrefix = "arXiv",
    doi = "10.1103/PhysRevD.60.123507",
    journal = "Phys. Rev. D",
    volume = "60",
    pages = "123507",
    year = "1999"
}

@article{Feldbrugge:2017fcc,
    author = "Feldbrugge, Job and Lehners, Jean-Luc and Turok, Neil",
    title = "{No smooth beginning for spacetime}",
    eprint = "1705.00192",
    archivePrefix = "arXiv",
    primaryClass = "hep-th",
    doi = "10.1103/PhysRevLett.119.171301",
    journal = "Phys. Rev. Lett.",
    volume = "119",
    number = "17",
    pages = "171301",
    year = "2017"
}

@article{Brown:2017wpl,
    author = "Brown, Jon and Cole, Alex and Shiu, Gary and Cottrell, William",
    title = "{Gravitational decoupling and the Picard-Lefschetz approach}",
    eprint = "1710.04737",
    archivePrefix = "arXiv",
    primaryClass = "hep-th",
    reportNumber = "MAD-TH-17-06",
    doi = "10.1103/PhysRevD.97.025002",
    journal = "Phys. Rev. D",
    volume = "97",
    number = "2",
    pages = "025002",
    year = "2018"
}

@article{Feldbrugge:2017mbc,
    author = "Feldbrugge, Job and Lehners, Jean-Luc and Turok, Neil",
    title = "{No rescue for the no boundary proposal: Pointers to the future of quantum cosmology}",
    eprint = "1708.05104",
    archivePrefix = "arXiv",
    primaryClass = "hep-th",
    doi = "10.1103/PhysRevD.97.023509",
    journal = "Phys. Rev. D",
    volume = "97",
    number = "2",
    pages = "023509",
    year = "2018"
}

@article{Krishna:2026rma,
    author = "Krishna, Vineeth and Larsen, Finn",
    title = "{Allowable Complex Black Holes in the Euclidean Gravitational Path Integral}",
    eprint = "2602.05979",
    archivePrefix = "arXiv",
    primaryClass = "hep-th",
    reportNumber = "LITP-26-04",
    month = "2",
    year = "2026"
}

@article{Prestidge:1999uq,
    author = "Prestidge, Tim",
    title = "{Dynamic and thermodynamic stability and negative modes in Schwarzschild-anti-de Sitter}",
    eprint = "hep-th/9907163",
    archivePrefix = "arXiv",
    reportNumber = "DAMTP-1999-89",
    doi = "10.1103/PhysRevD.61.084002",
    journal = "Phys. Rev. D",
    volume = "61",
    pages = "084002",
    year = "2000"
}

@article{Dong:2019piw,
    author = "Dong, Xi and Marolf, Donald",
    title = "{One-loop universality of holographic codes}",
    eprint = "1910.06329",
    archivePrefix = "arXiv",
    primaryClass = "hep-th",
    doi = "10.1007/JHEP03(2020)191",
    journal = "JHEP",
    volume = "03",
    pages = "191",
    year = "2020"
}

@article{Kol:2006ga,
    author = "Kol, Barak",
    title = "{The Power of Action: The Derivation of the Black Hole Negative Mode}",
    eprint = "hep-th/0608001",
    archivePrefix = "arXiv",
    doi = "10.1103/PhysRevD.77.044039",
    journal = "Phys. Rev. D",
    volume = "77",
    pages = "044039",
    year = "2008"
}

@article{Garriga:1997wz,
    author = "Garriga, Jaume and Montes, Xavier and Sasaki, Misao and Tanaka, Takahiro",
    title = "{Canonical quantization of cosmological perturbations in the one-bubble open universe}",
    eprint = "astro-ph/9706229",
    archivePrefix = "arXiv",
    reportNumber = "UAB-FT-418, OU-TAP-63",
    doi = "10.1016/S0550-3213(97)00780-3",
    journal = "Nucl. Phys. B",
    volume = "513",
    pages = "343--374",
    year = "1998",
    note = "[Erratum: Nucl.Phys.B 551, 511--511 (1999)]"
}

@article{Horowitz:2025zpx,
    author = "Horowitz, Gary T. and Marolf, Donald and Santos, Jorge E.",
    title = "{Constraints are not enough}",
    eprint = "2505.13600",
    archivePrefix = "arXiv",
    primaryClass = "hep-th",
    doi = "10.1007/JHEP10(2025)031",
    journal = "JHEP",
    volume = "10",
    pages = "031",
    year = "2025"
}

@article{Kontsevich:2021dmb,
    author = "Kontsevich, Maxim and Segal, Graeme",
    title = "{Wick Rotation and the Positivity of Energy in Quantum Field Theory}",
    eprint = "2105.10161",
    archivePrefix = "arXiv",
    primaryClass = "hep-th",
    doi = "10.1093/qmath/haab027",
    journal = "Quart. J. Math. Oxford Ser.",
    volume = "72",
    number = "1-2",
    pages = "673--699",
    year = "2021"
}

@article{Witten:2021nzp,
    author = "Witten, Edward",
    title = "{A Note On Complex Spacetime Metrics}",
    eprint = "2111.06514",
    archivePrefix = "arXiv",
    primaryClass = "hep-th",
    month = "11",
    year = "2021"
}

@article{BenettiGenolini:2026raa,
    author = "Benetti Genolini, Pietro and Janssen, Oliver and Murthy, Sameer",
    title = "{Allowable complex metrics and the gravitational index of AdS$_5$ black holes}",
    eprint = "2601.23197",
    archivePrefix = "arXiv",
    primaryClass = "hep-th",
    month = "1",
    year = "2026"
}

@article{BenettiGenolini:2025jwe,
    author = "Benetti Genolini, Pietro and Murthy, Sameer",
    title = "{The gravitational index and allowable complex metrics}",
    eprint = "2503.20866",
    archivePrefix = "arXiv",
    primaryClass = "hep-th",
    doi = "10.1088/1751-8121/add7a7",
    journal = "J. Phys. A",
    volume = "58",
    number = "21",
    pages = "215401",
    year = "2025"
}

@article{Louko:1995jw,
    author = "Louko, Jorma and Sorkin, Rafael D.",
    title = "{Complex actions in two-dimensional topology change}",
    eprint = "gr-qc/9511023",
    archivePrefix = "arXiv",
    reportNumber = "SU-GP-95-5-1, WISC-MILW-95-TH-16, MDDP-PP-96-40",
    doi = "10.1088/0264-9381/14/1/018",
    journal = "Class. Quant. Grav.",
    volume = "14",
    pages = "179--204",
    year = "1997"
}

@article{Chakravarty:2024bna,
    author = "Chakravarty, Joydeep and Maloney, Alexander and Namjou, Keivan and Ross, Simon F.",
    title = "{A new observable for holographic cosmology}",
    eprint = "2407.04781",
    archivePrefix = "arXiv",
    primaryClass = "hep-th",
    doi = "10.1007/JHEP10(2024)184",
    journal = "JHEP",
    volume = "10",
    pages = "184",
    year = "2024"
}

@article{Kolanowski:2026aaa,
    author = "Kolanowski, Maciej and Marolf, Donald and  Zheng, Wenwen and  Wang, Zi-Yue",
    title = "{Superconformal indices and black hole saddles}",
    journal="(in preparation)",
    year="2026"
}

@article{Chen:2023mbc,
    author = "Chen, Yiming and Turiaci, Gustavo J.",
    title = "{Spin-statistics for black hole microstates}",
    eprint = "2309.03478",
    archivePrefix = "arXiv",
    primaryClass = "hep-th",
    doi = "10.1007/JHEP04(2024)135",
    journal = "JHEP",
    volume = "04",
    pages = "135",
    year = "2024"
}

@article{Grabovsky:2024vnb,
    author = "Grabovsky, David and Kolanowski, Maciej",
    title = "{Spin-refined partition functions and $ \mathcal{CRT} $ black holes}",
    eprint = "2406.07609",
    archivePrefix = "arXiv",
    primaryClass = "hep-th",
    doi = "10.1007/JHEP12(2024)013",
    journal = "JHEP",
    volume = "12",
    pages = "013",
    year = "2024"
}

@article{Iliesiu:2026,
    author = "C. Goker and L.V. Iliesiu and E. Tabor",
    title ="A note on the quantization of angular momentum for black holes",
    journal="in preparation",
    year = "2026"
}

@article{Maloney:2007ud,
    author = "Maloney, Alexander and Witten, Edward",
    title = "{Quantum Gravity Partition Functions in Three Dimensions}",
    eprint = "0712.0155",
    archivePrefix = "arXiv",
    primaryClass = "hep-th",
    doi = "10.1007/JHEP02(2010)029",
    journal = "JHEP",
    volume = "02",
    pages = "029",
    year = "2010"
}

@article{Turiaci:2025xwi,
    author = "Turiaci, Gustavo J. and Wu, Chih-Hung",
    title = "{The wavefunction of a quantum S$^{1}$ {\texttimes} S$^{2}$ universe}",
    eprint = "2503.14639",
    archivePrefix = "arXiv",
    primaryClass = "hep-th",
    doi = "10.1007/JHEP07(2025)158",
    journal = "JHEP",
    volume = "07",
    pages = "158",
    year = "2025"
}
\end{document}